\documentclass[journal,12pt,onecolumn,draftclsnofoot]{IEEEtran}
\usepackage{algorithm,algorithmicx}

\makeatletter

\usepackage{url}
\newcommand\fs@betterruled{%
  \def\@fs@cfont{\bfseries}\let\@fs@capt\floatc@ruled
  \def\@fs@pre{\vspace*{5pt}\hrule height.8pt depth0pt \kern2pt}%
  \def\@fs@post{\kern2pt\hrule\relax}%
  \def\@fs@mid{\kern2pt\hrule\kern2pt}%
  \let\@fs@iftopcapt\iftrue}
\floatstyle{betterruled}
\restylefloat{algorithm}
\makeatother

\newcommand{\squeezeup}{\vspace{-2mm}}
\usepackage{comment}
\usepackage{cite}
\usepackage{amsmath,amssymb,amsfonts}
\usepackage{algpseudocode}
\usepackage{graphicx}
\usepackage{textcomp}
\usepackage{xcolor}
\usepackage{tabularx,booktabs}
\newcolumntype{C}{>{\centering\arraybackslash}X}
\DeclareFontFamily{OT1}{pzc}{}
\DeclareFontShape{OT1}{pzc}{m}{it}{<-> s * [1.3] pzcmi7t}{}
\DeclareMathAlphabet{\mathpzc}{OT1}{pzc}{m}{it}
\DeclareMathAlphabet\mathbfcal{OMS}{cmsy}{b}{n}

\renewcommand{\d}{\mathbf{d}}

\renewcommand{\t}{\mathbf{t}}

\newcommand{\w}{\mathbf{w}}
\newcommand{\x}{\mathbf{x}}
\newcommand{\y}{\mathbf{y}}
\newcommand{\z}{\mathbf{z}}

\newcommand{\C}{\mathbf{C}}

\newcommand{\G}{\mathbf{G}}
\renewcommand{\H}{\mathbf{H}}
\newcommand{\I}{\mathbf{I}}

\renewcommand{\L}{\mathbf{L}}

\newcommand{\U}{\mathbf{U}}
\newcommand{\V}{\mathbf{V}}
\newcommand{\W}{\mathbf{W}}
\newcommand{\X}{\mathbf{X}}
\newcommand{\Y}{\mathbf{Y}}

\newcommand{\argmin}{\operatornamewithlimits{argmin}}

\begin{document}

\title{Joint Analog and Digital Transceiver Design for Wideband Full Duplex MIMO Systems}

\author{Md Atiqul Islam,~\IEEEmembership{Student Member,~IEEE,}
        George C. Alexandropoulos,~\IEEEmembership{Senior Member,~IEEE,}
        and~Besma Smida,~\IEEEmembership{Senior Member,~IEEE}% <-this % stops a space
\thanks{Part of this work has been presented in \textit{IEEE PIMRC}, Istanbul, Turkey, Sep. 2019 \cite{islam2019unified}.}
\thanks{M.~A.~Islam and B. Smida are with the Department of Electrical and Computer Engineering, University of Illinois at Chicago, USA. (e-mails: \{mislam23,smida\}@uic.edu)}
\thanks{G.~C.~Alexandropoulos is with the Department of Informatics and Telecommunications, National and Kapodistrian University of Athens, Panepistimiopolis Ilissia, 15784 Athens, Greece. (e-mail: alexandg@di.uoa.gr)}
}

% The paper headers
% \markboth{IEEE TRANSACTIONS ON COMMUNICATIONS}%
% {Shell \MakeLowercase{\textit{et al.}}: Bare Demo of IEEEtran.cls for IEEE Journals}
\maketitle

\begin{abstract}
In this paper, we propose a wideband Full Duplex (FD) Multiple-Input Multiple-Output (MIMO) communication system comprising of an FD MIMO node simultaneously communicating with two multi-antenna UpLink (UL) and DownLink (DL) nodes utilizing the same time and frequency resources. To suppress the strong Self-Interference (SI) signal due to simultaneous transmission and reception in FD MIMO systems, we propose a joint design of Analog and Digital (A/D) cancellation as well as transmit and receive beamforming capitalizing on baseband Orthogonal Frequency-Division Multiplexing (OFDM) signal modeling. Considering practical transmitter impairments, we present a multi-tap wideband analog canceller architecture whose number of taps does not scale with the number of transceiver antennas and multipath SI components. We also propose a novel adaptive digital cancellation based on truncated singular value decomposition that reduces the residual SI signal estimation parameters. To maximize the FD sum rate, a joint optimization framework is presented for A/D cancellation and digital beamforming. Finally, our extensive waveform simulation results demonstrate that the proposed wideband FD MIMO design exhibits higher SI cancellation capability with reduced complexity compared to existing cancellation techniques, resulting in improved achievable rate performance.
\end{abstract}

\begin{IEEEkeywords}
Analog and digital cancellation, beamforming, full duplex, impairments, IQ imbalance, nonlinear amplification, wideband MIMO, optimization, self-interference cancellation.
\end{IEEEkeywords}

\IEEEpeerreviewmaketitle

\section{Introduction}
\IEEEPARstart{F}{uture} wireless communication systems need to accommodate the explosive growth in data traffic demand through efficient utilization of limited frequency spectrum. Recent advances in Full Duplex (FD) communication technology demonstrate the potential of a substantial spectral efficiency improvement over the conventional frequency- and time-division duplexing systems through simultaneous UpLink (UL) and DownLink (DL) communication in the same frequency and time resources \cite{islam2019unified,Samsung,sabharwal2014band,kolodziej2019band,smida2017reflectfx,korpi2016full}. The exploitation of wideband Multiple-Input Multiple-Output (MIMO) systems provides further spectral performance boost due to enhanced spatial degrees of freedom (DoF) offered by the plurality of Transmitter (TX) and Receiver (RX) antennas and larger bandwidths \cite{riihonen2011mitigation,everett2016softnull,alexandropoulos2017joint,bharadia2014full,masmoudi2017channel,anttila2014modeling,bharadia2013full}. In addition, FD MIMO radios have recently been considered for certain physical-layer-based latency improvement through simultaneous communication of data and control signals \cite{islam2020simultaneous,mirza2018performance,Islam_2020_Sim_Multi}. Thus, enabling FD in conjunction with wideband MIMO operation can meet the stringent throughput and latency requirements of beyond 5th Generation (5G) wireless communication systems with limited spectrum resources \cite{Samsung}.

The simultaneous transmission and reception in wideband FD systems induce strong in-band Self-Interference (SI) signals at the FD receivers due to the inevitable limited isolation between the TX and RX blocks \cite{sabharwal2014band}.
To suppress the strong SI signal, first, analog cancellation is employed at the input of the RX blocks at the FD node to ensure that none of the reception Radio Frequency (RF) components (i.e., Low Noise Amplifiers (LNAs), In-phase Quadrature (IQ) mixers) goes into saturation due to high SI power, while ensuring that the dynamic range of the Analog-to-Digital-Converters (ADCs) is large enough to capture the residual SI and the naturally weak desired signal with sufficient precision \cite{korpi2016full}. Compared to a narrowband FD Single-Input Single-Output (SISO) system, where a single direct SI coupling path exists between TX and RX, the wideband FD MIMO analog cancellation design is much more challenging. This happens because each RX chain suffers from the direct SI signals introduced by all the TX antennas as well as their multipath SI components created by environmental reflections \cite{bharadia2014full,khaledian2018inherent,kolodziej2016multitap,khaledian2018robust,korpi2016full,chen2019wideband,antonio2013adaptive,lee2019analysis,le2020beam,huberman2014mimo,duarte2020full,cao2020integrated}. For an $N_{\rm TX}\times N_{\rm RX}$ FD MIMO transceiver, the narrowband analog canceller requires $N_{\rm TX}N_{\rm RX}$ taps  to suppress the direct SI coupling paths, where each cancellation tap includes time delays, tunable bandpass filters, phase shifters, and attenuators. Considering a wideband communication with $L$ multipath SI components being strong enough to run the RX RF chains into saturation, the same FD MIMO system would require an appropriate wideband analog canceller with $N_{\rm TX}N_{\rm RX}L$ taps; such analog cancellers are hereinafter referred to as \textit{full-tap} cancellers.

After the analog cancellation, digital domain SI mitigation techniques are applied at the RX baseband to suppress the residual SI signal below the noise floor, which is still large enough to overwhelm the weak desired signal \cite{sabharwal2014band,kolodziej2019band}. Digital cancellation is accomplished by reconstructing and reciprocally combining the residual SI signal at the FD RX through extensive SI channel modeling and
exploiting the fact that each FD node has knowledge of its ideal transmit signal in the digital domain \cite{riihonen2011mitigation,bharadia2013full}. Since the residual SI signal is impacted by TX hardware impairments, an appropriate SI channel model must include the SI coupling paths and the nonlinear distortions induced by the transceiver chain's practical RF components, specifically the image effect due to the gain and phase imbalance of the IQ mixer and Power Amplifier (PA) nonlinearities. For single-antenna FD systems, baseband modeling of these nonlinear distortions has been performed to provide appropriate digital cancellation \cite{bharadia2013full,ahmed2015all,korpi2014widely,islam2019comprehensive}. Akin to the analog canceller, the wideband FD MIMO operation increases the computational complexity of the digital cancellation since the number of linear and nonlinear components to be estimated increases with the number of TX/RX chains and SI channel paths. Moreover, signal modeling in wideband FD MIMO systems requires multi-carrier designs (i.e.~Orthogonal Frequency-Division Multiplexing (OFDM)), as the channel becomes frequency selective due to larger bandwidth. 

\subsection{Related works on FD MIMO SI Cancellation}
For narrowband or frequency-flat FD MIMO systems, full-tap analog cancellers connecting all TX outputs to RX inputs are usually employed, where the number of taps increases with the number of TX/RX RF chains \cite{riihonen2011mitigation,masmoudi2017channel}. To reduce this hardware complexity, analog SI canceller designs exploiting AUXiliary (AUX) TX structures and/or joint design of TX/RX beamformers were studied in \cite{huberman2014mimo,duarte2020full,george2018journal}, where the analog cancellation signals were injected into each of the RX inputs using separate TX RF chains. Although these techniques reduce the analog canceller hardware complexity, they are unable to suppress the nonlinear SI components due to the non-ideal RF front-end hardware rendering the RX chains into saturation \cite{kolodziej2019band}. In our previous work \cite{islam2019unified}, we presented a unified low complexity Analog and Digital (A/D) cancellation for narrowband FD MIMO systems.
For wideband FD MIMO systems in \cite{antonio2013adaptive,bharadia2014full}, full-tap analog cancellers with adaptive filters were utilized to provide sufficient analog SI cancellation. In \cite{lee2019analysis}, a full-tap wideband FD MIMO RF canceller was presented with a tunable time delay circuit, which employed reflected type phase shifters to emulate the true time delays of the SI channel. Recently, a full-tap beam-based RF cancellation approach was introduced in \cite{le2020beam}, which employed analog Least Mean-Squared (LMS) loops as the adaptive filters for SI mitigation in FD massive MIMO systems. Those analog LMS loops include time delay generators, down-converters, Low-Pass Filters (LPFs), and up-converters. An integrated LMS adaptive wideband FD MIMO RF canceller was proposed in \cite{cao2020integrated}, where the time delay of the cancellation was generated using an $N$-path filter.
However, the hardware complexity of all the above full-tap RF cancellers scales with the number of TX and RX RF chains as well as the number of SI multipath components, rendering the practical implementation of the analog SI cancellation unit a core design bottleneck.

Alleviating the need for analog SI cancellation, spatial suppression techniques were presented in \cite{riihonen2011mitigation,everett2016softnull} for narrowband FD MIMO systems, where the SI suppression was solely handled by the digital TX/RX beamformers. However, those spatial suppression techniques were unable to cancel the SI in high TX power and often resulted in reductions of the data rates for both the UL and DL signals of interest. This stemmed from the fact that some of the available spatial DoFs were devoted to mitigating SI \cite{alexandropoulos2017joint}. To avoid such issues, digital cancellation techniques exploiting SI signal modeling were utilized in practice to supplement the analog canceller in suppressing the SI signal. To achieve sufficient SI suppression, existing digital cancellation approaches capitalize on models for the PA impairments \cite{bharadia2014full} and IQ mixer image effect \cite{korpi2014widely}, or rely on cascaded SI designs taking into account both nonidealities \cite{anttila2014modeling}. However, the number of estimation parameters of those models grows with the number of TX/RX RF chains and SI channel components.
 To reduce the number of parameters for the FD MIMO system, a digital canceller based on Principle Component Analysis (PCA) was provided in \cite{korpi2017nonlinear}. Furthermore, in \cite{ng2012dynamic,taghizadeh2018hardware,radhakrishnan2021hardware}, the authors considered FD MIMO OFDM signal modeling to design rate maximizing TX/RX beamformers. However, these techniques assumed full-tap RF cancellers to achieve certain SI suppression levels. 

\subsection{Contributions}
In this paper, we present a joint A/D SI cancellation with TX/RX beamforming approach for wideband FD MIMO systems considering the effect of non-linear hardware distortions and multipath SI components.
The main contributions of this paper are summarized as follows:
 \begin{itemize}
     \item We propose a novel joint wideband analog SI cancellation and TX/RX beamforming approach for multi-user FD MIMO systems in the presence of TX RF chain impairments, where the multipath SI components are suppressed using reduced analog cancellations taps compared to existing FD MIMO solutions.
     \item A comprehensive OFDM signal modeling of the proposed FD MIMO system is derived, including baseband equivalent models of the TX RF chain impairments, corresponding wideband channels, and A/D SI cancellers.
     \item We present a novel adaptive digital canceller based on the Truncated Singular Value Decomposition (TSVD) that reduces the computational complexity of conventional digital SI cancellation while successfully suppressing the residual SI signal after analog cancellation below the RX noise floor.
     \item A joint optimization framework for A/D cancellation and TX/RX beamforming is presented to maximize the achievable sum-rate performance of the considered three-node wideband FD MIMO OFDM system.
     \item Finally, we perform extensive waveform simulations to illustrate the proposed A/D SI cancellation performance and provide comparisons with the relevant state-of-the-art methods. It is demonstrated that our proposed wideband canceller in conjunction with TX/RX beamforming exhibits superior SI mitigation capability with reduced complexity (less than $50\%$ analog taps) compared to the existing full-tap cancellers in the presence of TX hardware impairments.
 \end{itemize}
\subsection{Organization and Notations}
The rest of the paper is organized as follows. In Section~\ref{sec: system_signal}, the baseband signal model of the considered wideband FD MIMO OFDM system is presented. Then, in Sections~\ref{sec: analog_canceller} and \ref{sec: digital}, we present the proposed A/D SI canceller alongside the joint optimization framework. Section~\ref{sec: Simulation} includes the performance evaluations of the proposed SI cancellation approach via extensive waveform simulations. Finally, the conclusions are drawn in Section~\ref{sec: conclusion}.

{
\textit{Notations:} Vectors and matrices are denoted by boldface lowercase and boldface capital letters, respectively. The transpose, Hermitian transpose, and conjugate of $\mathbf{A}$ are denoted by $\mathbf{A}^{\rm T}$, $\mathbf{A}^{\rm H}$, and $\mathbf{A}^*$, respectively, and $\det(\mathbf{A})$ is $\mathbf{A}$'s determinant, while $\mathbf{I}_{n}$ ($n\geq2$) is the $n\times n$ identity matrix. $\|\mathbf{a}\|$ stands for the Euclidean norm of $\mathbf{a}$, $\mathbf{a}^{\circ n}$ denotes the Hadamard power operation to the factor $n$, operand $\odot$ represents the Hadamard entry-wise product, ${\rm col}\{\mathbf{a}_1,\mathbf{a}_2,\ldots,\mathbf{a}_n\}$ is a column vector resulting after vertically concatenating vectors $\mathbf{a}_1,\mathbf{a}_2,\ldots,\mathbf{a}_n$, and ${\rm diag}\{\mathbf{a}\}$ denotes a square diagonal matrix with $\mathbf{a}$'s elements in its main diagonal. $[\mathbf{A}]_{i,j}$, $[\mathbf{A}]_{(i,:)}$, and $[\mathbf{A}]_{(:,j)}$ represent $\mathbf{A}$'s $(i,j)$-th element, $i$-th row, and $j$-th column, respectively, while $[\mathbf{a}]_{i}$ denotes the $i$-th element of $\mathbf{a}$. $\mathbb{C}$ represents the complex number set, $\mathbb{E}\{\cdot\}$ is the expectation operator, and $|\cdot|$ denotes the amplitude of a complex number. The rest of the notations used throughout this paper are listed in Table~\ref{tab: variables}.}

\begin{table*}[!tpb]
    \caption{{The notations of this paper.}}
    \label{tab: variables}
    \begin{tabularx}{0.48\textwidth}{l l}
        \toprule
         Variable & Definition \\
        \midrule
         $N_{\text{TX},b}$ & Number of TX antennas at node $b$\\
         $N_{\text{RX},b}$ & Number of RX antennas at node $b$\\
         $N_{\text{RX},m_1}$ & Number of RX antennas at node $m_1$\\
         $N_{\text{TX},m_2}$ & Number of TX antennas at node $m_2$\\
         $N_{c}$ & Number of subcarriers\\
         $n$ & Subcarrier index\\
         $d_b$ & Number of data streams at node $b$\\
         $d_{m_2}$ & Number of data streams at node $m_2$\\
         $\mathpzc{s}_{b,n}$ & $n$-th subcarrier symbol vector at node $b$\\
         $\mathbfcal{V}_{b,n}$ & $n$-th subcarrier TX Beamformer at node $b$\\
         $\mathpzc{s}_{m_2,n}$ & $n$-th subcarrier symbol vector at node $m_2$\\
         $\mathbfcal{V}_{m_2,n}$ & $n$-th subcarrier TX Beamformer at node $m_2$\\
         $\x_{b}[k]$ & Node $b$ symbol vector at time $k$\\
         $\x_{m_2}[k]$ & Node $m_2$ symbol vector at time $k$\\
         $\G_{1,b}$ & Linear power allocation matrix at node $b$\\
         $\G_{1,m_2}$ & Linear power allocation matrix at node $m_2$\\
         $\z_{b}[k]$ & Nonlinear TX signal at node $b$\\
         $\z_{m_2}[k]$ & Nonlinear TX signal at node $m_2$\\
         $\widetilde{\x}_b[k]$ & Node $b$ TX output at time instant $k$\\
         $\widetilde{\x}_{m_2}[k]$ & Node $m_2$ TX output at time instant $k$\\
         $\H_{\rm DL}[\ell]$ & Wideband DL channel between node $b$ and $m_1$\\
         $L_{\rm DL}$ & DL channel delay taps\\
         $\H_{\rm UL}[\ell]$ & Wideband UL channel between node $m_2$ and $b$\\
         $L_{\rm UL}$ & UL channel delay taps\\
         $\H_{\rm SI}[\ell]$ & Wideband SI channel\\
         $L_{\rm SI}$ & SI channel delay taps\\
         $\C_{b}[\ell]$ & Coefficients of the analog SI canceller\\
         $L_{\rm C}$ & Delay taps of analog SI canceller\\
         $\y_{m_{1}}[k]$ & Baseband received signal at node $m_1$\\
        \bottomrule
    \end{tabularx}
    \begin{tabularx}{0.48\textwidth}{l l}
        \toprule
         Variable & Definition \\
        \midrule
         $\y_{b}[k]$ & Baseband received signal at node $b$\\
         $\w_{m_1}[k]$ & AWGN vector at node $m_1$\\
         $\w_{b}[k]$ & AWGN vector at node $b$\\
         $\d_{b}[k]$ & Digital SI cancellation signal at node $b$\\
         $\mathpzc{r}_{b,n}$ & $n$-th subcarrier FFT output at node $b$\\
         $\widehat{\mathpzc{s}}_{b,n}$ & Estimated symbol vector of ${\mathpzc{s}}_{b,n}$\\
         $\widehat{\mathpzc{s}}_{m_2,n}$ & Estimated symbol vector of ${\mathpzc{s}}_{m_2,n}$\\
         $\mathbfcal{U}_{m_1,n}$ & $n$-th subcarrier RX beamformer at node $m_1$\\
         $\mathbfcal{U}_{b,n}$ & $n$-th subcarrier RX beamformer at node $b$\\
         $\mathbfcal{H}_{{\rm DL},n}$ & Frequency representation of DL channel\\
         $\mathbfcal{H}_{{\rm UL},n}$ & Frequency representation of UL channel\\
         $\mathbfcal{H}_{{\rm SI},n}$ & Frequency representation of SI channel\\
         $\mathbfcal{C}_{b,n}$ & Frequency representation of analog SI canceller\\
         $\mathpzc{z}_{b,n}$ & Frequency representation of $\z_{b}[k]$\\
         $\mathpzc{z}_{m_2,n}$ & Frequency representation of $\z_{m_2}[k]$\\
         $\mathpzc{w}_{b,n}$ & Frequency representation of $\w_{b}[k]$\\
         $\mathpzc{w}_{m_1,n}$ & Frequency representation of $\w_{m_1}[k]$\\
         $\mathpzc{d}_{b,n}$ & Frequency representation of $\d_{b}[k]$\\
         $\L_1[\ell]$ & Analog SI canceller MUX configurations\\
         $\L_2[\ell]$ & Analog SI canceller coefficients of $\ell$-th tap\\
         $\L_3[\ell]$ & Analog SI canceller DEMUX configurations\\
         $\widehat{\mathcal{R}}_{\text{UL},n}$ & Achievable UL rate of $n$-th subcarrier\\
         $\widehat{\mathcal{R}}_{\text{DL},n}$ & Achievable DL rate of $n$-th subcarrier\\
         $\widehat{\mathbfcal{Q}}_{b,n}$ & Estimated IpN covariance matrix at node $b$\\
         $\widehat{\mathbfcal{Q}}_{m_1,n}$ & Estimated IpN covariance matrix at node $m_1$\\
         ${\rm P}_{b}$ & Maximum transmit power at node $b$\\
         ${\rm P}_{m_2}$ & Maximum transmit power at node $m_2$\\
         ${\lambda}_{b}$ & RF chain saturation threshold at node $b$\\
         &\\
        \bottomrule
    \end{tabularx}
\end{table*}
 
\section{System and Signal Models}\label{sec: system_signal}
\begin{figure*}[!t]
\centering
\includegraphics[width=1\textwidth]{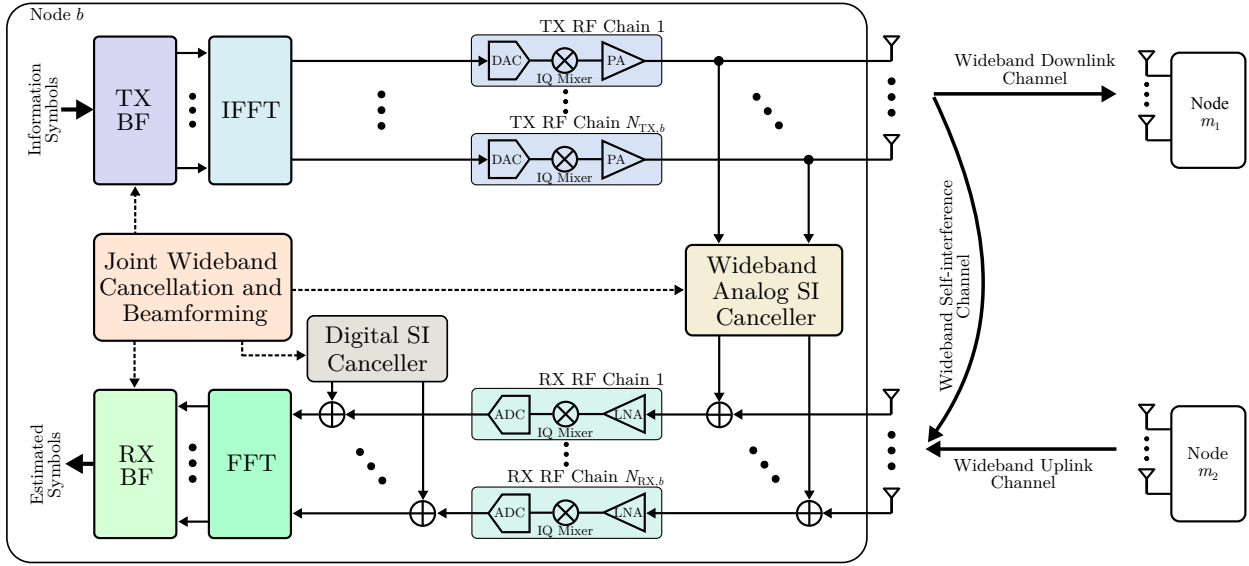}
\caption{The considered three-node communication system and the proposed wideband FD MIMO architecture. The FD MIMO node $b$ incorporates processing blocks for analog SI cancellation and digital TX/RX beamforming similar to \cite{alexandropoulos2017joint}, as well as for a digital cancellation block for treating the residual SI at the output of the TX RF chains. All these blocks are jointly optimized, realizing the desired wideband FD operation in a hardware-efficient way. The HD multi-antenna nodes $m_1$ and $m_2$ communicate with node $b$ in the downlink and uplink directions, respectively.}
\label{fig: transceiver}
% \squeezeup
% \squeezeup
\squeezeup
\end{figure*}

We consider the three-node FD wireless communication system of Fig$.$~\ref{fig: transceiver} comprising of an FD MIMO Base Station (BS) node $b$ communicating concurrently with two Half Duplex (HD) multi-antenna nodes: node $m_1$ in the downlink and node $m_2$ in the uplink direction. The FD MIMO node $b$ is assumed to be equipped with $N_{\text{TX},b}$ TX and $N_{\text{RX},b}$ RX antenna elements. Each TX antenna is attached to a dedicated RF chain that consists of a Digital to Analog Converter (DAC), IQ mixer, and PA; similarly holds for the RX antennas and their attached RF chains, each containing LNA, IQ mixer, and ADC. The HD multi-antenna nodes $m_1$ and $m_2$ are assumed to have $N_{\text{RX},m_1}$ and $N_{\text{TX},m_2}$ antennas, respectively, with each of their antennas connected to a dedicated RF chain. All three nodes are considered capable of performing digital beamforming and OFDM operations with $N_c$ subcarriers.
\squeezeup
\squeezeup
\subsection{Downlink TX and RX Signal Modeling}
During DL transmission, the FD MIMO node $b$ sends $d_b\leq \min\{N_{\text{TX},b},N_{\text{RX},m_1}\}$ data streams multiplexed at each subcarrier $n = \{0,1,\ldots,N_c-1\}$ to the HD node $m_1$. 
The unit power symbol vector of $n$th subacarrier is denoted as $\mathpzc{s}_{b,n} \in \mathbb{C}^{d_b \times 1}$, which, in practice, is selected from a discrete modulation set. The symbol vector $\mathpzc{s}_{b,n}$ is linearly precoded by the TX beamformer $\mathbfcal{V}_{b,n}\in \mathbb{C}^{N_{\text{TX},b} \times d_b}$. Without loss of generality, we assume that $\mathbfcal{V}_{b,n}$ have unit norm columns. The precoded symbols are converted to time-domain samples using the Inverse Fast Fourier Transform (IFFT) operation. To prevent Inter-Symbol Interference (ISI), a cyclic prefix is appended in front of each IFFT block; in the subsequent analysis, we ignore the cyclic prefix for simplicity. The output of the node $b$ TX baseband block after the IFFT operation at a discrete time instant $k$ is represented as
\begin{equation}\label{eq: ifft_b}
        \x_{b}[k] \triangleq \frac{1}{\sqrt{N_c}} \sum\limits_{n=0}^{N_c-1} \mathbfcal{V}_{b,n}\mathpzc{s}_{b,n} e^{\frac{j2\pi n k}{N_c}}.
\end{equation}

\textbf{Baseband Modeling of TX RF Chain Impairments:} As shown in Fig$.$~\ref{fig: transceiver}, the node $b$ baseband samples in $\x_{b}[k]$ are fed to the TX RF chains for upconversion and amplification. We introduce a baseband equivalent model for each of these RF chains incorporating IQ imbalances and PA nonlinearities \cite{korpi2014widely}, assuming that the RF chains are identical. 
Upon entering the TX RF chain of node $b$, each baseband sample goes through the IQ mixer for upconversion to the carrier frequency. In practical IQ mixers, a mirror image of the original signal with certain image attenuation is induced by the IQ phase and gain imbalances. Denoting the input at the $i$th ($i=1,2,\ldots,N_{\text{TX},b}$) TX RF chain of node $b$ at time instant $k$ as $[\x_b[k]]_{i}$, the IQ mixer output can be written as \cite[eq. (8)]{korpi2014widely}
\begin{equation}\label{eq: IQ_b}
    \begin{split}
        [\x_b[k]]_{i}^{\rm IQ} \triangleq \mu_{1}[\x_b[k]]_{i} + \mu_{2}[\x_b[k]]_{i}^{*},
    \end{split}
\end{equation}
where $\mu_{1}\triangleq(1+g e^{-j\theta})/2$ and $\mu_{2}\triangleq(1-g e^{j\theta})/2$ with $g$ and $\theta$ are representing the gain and phase imbalances, respectively. It is noted that the Image Rejection Ratio, defined as $\text{IRR}\triangleq\left|\mu_{1}/\mu_{2}\right|^2$, represents the strength of the IQ induced conjugate term\cite{korpi2014full}.

Before transmission, the upconverted signal is fed into the PA for amplification while satisfying the TX power constraint. Note that practical PAs exhibit varying degrees of nonlinearity. However, we consider a quasi memoryless PA model of third-order nonlinearity, as it is the most dominant distortion in practice, and all the even-power harmonics lie out of the band and will be cut off by the RF low pass filter at the RXs \cite{gu2005rf, korpi2014widely, zhou2005baseband}. For this PA model, the baseband equivalent of each $i$th PA output at time instant $k$ is given using \eqref{eq: IQ_b} as
\begin{equation}\label{eq: PA_output}
    \begin{split}
        [\x_b[k]]_{i}^{\rm PA} & \triangleq \sum_{p=1,3} \nu_p \left|[\x_b[k]]_{i}^{\rm IQ}\right|^{p-1} [\x_b[k]]_{i}^{\rm IQ} \\
        & =  g_{1,i}[\x_b[k]]_{i} + g_{2,i}[\x_b[k]]_{i}^* + g_{3,i}[\x_b[k]]_{i}^3 + g_{4,i}[\x_b[k]]_{i}^2[\x_b[k]]_{i}^*
        \\& \quad + g_{5,i}[\x_b[k]]_{i}([\x_b[k]]_{i}^*)^2 + g_{6,i}([\x_b[k]]_{i}^*)^3, 
    \end{split}
\end{equation}
where $p$ represents the nonlinearity order{\footnote{{Note that \eqref{eq: PA_output} is general enough to model various degrees of nonlinearities in the TX RF chains; this can be accomplished by setting $p$ to the desired nonlinearity order.}}} and the six gain components $g_{\ell,i}$ with $\ell=1,2,\ldots,6$ are derived as 
\begin{equation}   \label{eq: small_gs} 
    \begin{split}
        g_{1,i}&\triangleq\mu_{1}\nu_{1},\; g_{2,i} \triangleq \mu_{2}\nu_{1},\; g_{3,i}\triangleq \mu^2_{1}\mu^{*}_{2}\nu_{3},\;
        g_{4,i}\triangleq (2|\mu_{1}|^2\mu_{1}\!+\!|\mu_{2}|^2\mu_{1})\nu_{3},\\
        g_{5,i}&\triangleq (2|\mu_{1}|^2\mu_{2}\!+\!|\mu_{2}|^2\mu_{2})\nu_{3},\,g_{6,i}\triangleq\mu^{*}_{1}\mu^{2}_{2}\nu_{3},
    \end{split}
\end{equation}
where $\nu_{1}$ denotes the PA linear gain and $\nu_{3}\triangleq{\nu_{1}}{/}{{(\rm IIP3)}^2}$ is the gain of the third-order nonlinear distortions with ${\rm IIP3}$ representing the third-order Input-referred Intercept Point of the PA\cite{gu2005rf}.

Based on \eqref{eq: PA_output} and after some algebraic manipulations, the baseband representation of the impaired transmitted signal from the $N_{\text{TX},b}$ TX antennas of the FD MIMO node $b$ in the DL direction can be expressed as
\begin{equation}\label{eq: signal_DL}
        \widetilde{\x}_b[k] \triangleq  \G_{1,b}\x_b[k] + \z_{b}[k] = \G_{b}\boldsymbol{\psi}_{b}[k],
\end{equation}
where $\G_{1,b}\triangleq\text{diag}\{g_{1,1},g_{1,2},\dots,g_{1,N_{\text{TX},b}}\}$ is the power allocation matrix of the linear components of the TX signal and $\z_{b}[k]$ denotes its nonlinear part, which is given by
\begin{equation}\label{eq: signal_impairments}
\begin{split}
        \z_{b}[k] \triangleq &\; \G_{2,b}\x_b^*[k] + \G_{3,b}(\x_b[k])^{\circ 3} \\&+ \G_{4,b}(\x_b[k])^{\circ 2}\odot\x_b^*[k] + \G_{5,b}\x_b[k]\odot(\x_b^*[k])^{\circ 2}  +\G_{6,b}(\x_b^{*}[k])^{\circ 3}.
\end{split}        
\end{equation}
In the latter expression, $\G_{\ell,b}\triangleq\text{diag}\{g_{\ell,1},g_{\ell,2},\dots,g_{\ell,N_{\text{TX},b}}\}$ for $\ell=2,3,\ldots,6$ representing the coefficient matrices for the nonlinear components of $\widetilde{\x}_b[k]$ is defined in the similar way to $\G_{1,b}$. In \eqref{eq: signal_DL}, we also introduce the notation $\mathbf{G}_b\in\mathbb{C}^{N_{\text{TX},b}\times 6N_{\text{TX},b}}$ for the augmented power allocation matrix, and the vertically arranged signal vector $\boldsymbol{\psi}_b[k]\in\mathbb{C}^{6N_{\text{TX},b}\times 1}$ including the image and nonlinear components. The matrices are given by
\begin{equation}\label{eq:augment_signal}
    \begin{split}
        \mathbf{G}_b &\triangleq \left[\mathbf{G}_{1,b}\,\, \mathbf{G}_{2,b}\,\,\mathbf{G}_{3,b}\,\,\mathbf{G}_{4,b}\,\mathbf{G}_{5,b}\,\,\mathbf{G}_{6,b}\right],\\
        \boldsymbol{\psi}_b[k]&\triangleq\text{col}\{\x_b[k],\x_b^*[k],\x_b[k]^{\circ 3},\x_b[k]^{\circ 2}\odot\x_b^*[k],\x_b[k]\odot(\x_b^*[k])^{\circ 2},(\x_b^*[k])^{\circ 3}\}.
    \end{split}
\end{equation}
We finally make the practical assumption that the DL signal transmission is power limited to ${\rm P}_b$ such that $\mathbb{E}\{\|\G_{1,b}\x_b[k] +\z_{b}[k]\|^2\}\leq {\rm P}_b$.

\textbf{DL Received Signal Model:} The transmitted DL signal $\widetilde{\x}_b[k]$ is received at the HD RX node $m_1$ after passing through the wideband DL channel denoted by $\H_{\rm DL}[\ell]\in \mathbb{C}^{N_{\text{RX},m_1}\times N_{\text{TX},b}}$, $\forall \ell = \{0,1,\ldots,L_{\rm DL}-1\}$, where $L_{\rm DL}$ represents the number of DL channel paths. The received baseband signal $\y_{m_{1}}[k] \in \mathbb{C}^{N_{\text{RX},m_1}\times 1}$ of node $m_1$ at the discrete time instant $k$ is mathematically expressed as 
\begin{equation}\label{eq: received_m1}
    \begin{split}
        \y_{m_{1}}[k] &\triangleq \sum\limits_{\ell=0}^{L_{\rm DL}-1} \H_{\rm DL}[\ell]\widetilde{\x}_b[k-\ell] +  \w_{m_1}[k],
    \end{split}
\end{equation}
where $\w_{m_1}[k]$ represents the Additive White Gaussian Noise (AWGN) vector at node $m_1$ with covariance matrix $\sigma_{m_1}^2\I_{N_{\text{RX},m_1}}$. It is to be noted that we assume no inter-node interference between nodes $m_1$ and $m_2$ due to appropriate node scheduling\cite{alexandropoulos2016user,atzeni2016performance}. 

The baseband received signal $\y_{m_{1}}[k]$ is transformed to frequency domain using the Fast Fourier Transformation (FFT) operation, which is followed by RX beamforming for each subcarrier to obtain the estimated symbol vectors denoted by $\widehat{\mathpzc{s}}_{b,n}$, $\forall n = \{0,1,\ldots,N_c-1\}$. Denoting $n$th subcarrier RX beamformer as $\mathbfcal{U}_{m_1,n}\in \mathbb{C}^{d_b\times N_{\text{RX},m_1}}$, the linearly processed estimated symbol vector is written as
\begin{equation}\label{eq: est_s_b}
    \begin{split}
        \widehat{\mathpzc{s}}_{b,n} \triangleq &\; \mathbfcal{U}_{m_1,n} \left( \frac{1}{\sqrt{N_c}}\sum\limits_{k=0}^{N_c-1} \y_{m_{1}}[k]e^{-\frac{j2\pi kn}{N_c}}\right)\\
        =&\; \mathbfcal{U}_{m_1,n} \left( \frac{1}{\sqrt{N_c}}\sum\limits_{k=0}^{N_c-1}\left(\sum\limits_{\ell=0}^{L_{\rm DL}-1} \H_{\rm DL}[\ell]\widetilde{\x}_b[k-\ell] +  \w_{m_1}[k]\right)e^{-\frac{j2\pi kn}{N_c}}\right)\\
        =&\; \mathbfcal{U}_{m_1,n}\left(\mathbfcal{H}_{{\rm DL},n}\left(\G_{1,b}\mathbfcal{V}_{b,n}{\mathpzc{s}}_{b,n} + {\mathpzc{z}}_{b,n}\right) + \mathpzc{w}_{m_1,n}\right),
    \end{split}
\end{equation}
where $\mathbfcal{H}_{{\rm DL},n}\triangleq\sum\limits_{l=0}^{L_{\rm DL}-1} \H_{\rm DL}[l]e^{-\frac{j2\pi ln}{N_c}}$ is defined as $n$th subcarrier frequency domain representation of the wideband DL channel. Similarly, $\mathpzc{z}_{b,n}$ and $\mathpzc{w}_{m_1,n}$ represent the frequency transform of the TX nonlinear components vector $\z_{b}[k]$ and the AWGN vector at node $m_1$, respectively. The detailed derivation of \eqref{eq: est_s_b} is provided in Appendix A.

\subsection{Uplink TX and RX Signal Modeling}
Now, we model the UL signal transmitting from HD multi-antenna node $m_2$ to the FD MIMO node $b$. Similar to the node $b$ TX, the $n$th subcarrier symbol vector $\mathpzc{s}_{m_2,n} \in \mathbb{C}^{d_{m_2} \times 1}$ with $d_{m_2}\leq\min(N_{\text{TX},m_2},N_{\text{RX},b})$ is precoded by the unit norm TX beamformer $\mathbfcal{V}_{m_2,n}\in \mathbb{C}^{N_{\text{TX},m_2} \times d_{m_2}}$. The precoded symbol vectors are transformed to time domain samples $\x_{m_2}[k]\in \mathbb{C}^{N_{\text{TX},m_2} \times 1}$ using FFT operation identical to \eqref{eq: ifft_b}. The time domain samples are upconverted and amplified by the TX RF chains of node $m_2$ following the similar operation of node $b$ TX. Therefore, the TX output $\widetilde{\x}_{m_2}[k]\in \mathbb{C}^{N_{\text{TX},m_2} \times 1}$ at node $m_2$ is expressed as
\begin{equation}\label{eq: signal_UL}
        \widetilde{\x}_{m_2}[k] \triangleq  \G_{1,m_2}\x_{m_2}[k] + \z_{m_2}[k],
\end{equation}
where $\G_{1,m_2}$ is the power allocation matrix of the linear components of the TX signal and $\z_{m_2}[k]$ denotes its nonlinear part, defined similarly as \eqref{eq: signal_impairments}. The UL signal transmission is power limited to ${\rm P}_{m_2}$ such that $\mathbb{E}\{\|\G_{1,m_2}\x_{m_2}[k] +\z_{m_2}[k]\|^2\}\leq {\rm P}_{m_2}$.

\textbf{UL Received Signal Model:} The transmitted UL signal $\widetilde{\x}_{m_2}[k]$ is received at node $b$ after passing through the wideband UL channel denoted by $\H_{\rm UL}[\ell]\in \mathbb{C}^{N_{\text{RX},b}\times N_{\text{TX},m_1}}$, $\forall \ell = \{0,1,\ldots,L_{\rm UL}-1\}$, where $L_{\rm UL}$ represents the number of UL channel paths. Due to FD operation, the transmitted signal $\x_b[k]$ from node $b$ TX is also received at the node $b$ RX input after passing through the wideband SI channel denoted by $\H_{\rm SI}[\ell]\in \mathbb{C}^{N_{\text{RX},b}\times N_{\text{TX},b}}$, $\forall \ell = \{0,1,\ldots,L_{\rm SI}-1\}$, where $L_{\rm SI}$ is the number of SI channel delay taps. In addition to the UL and SI signals, the analog cancellation signal stemming from the output of the wideband analog SI canceller is fed into the RX inputs of node $b$, as shown in Fig$.$~\ref{fig: transceiver}. {Therefore, similar to \eqref{eq: received_m1}, the received signal $\y_b[k] \in \mathbb{C}^{N_{\text{RX},b}\times 1}$ is expressed as
\begin{equation}\label{eq: received_b}
        \y_b[k] \triangleq \sum\limits_{\ell=0}^{L_{\rm UL}-1} \H_{\rm UL}[\ell]\widetilde{\x}_m[k-\ell] + \sum\limits_{\ell'=0}^{L_{\rm SI}-1} \H_{\rm SI}[\ell']\widetilde{\x}_b[k-\ell'] + \sum\limits_{\ell''=0}^{L_{\rm C}-1} \C_{b}[\ell'']\widetilde{\x}_b[k-\ell''] + \w_b[k],
\end{equation}
where $\w_b[k]$ is the AWGN vector at this node with covariance matrix $\sigma_b^2\I_{N_{\text{RX},b}}$.} In this expression, $\C_{b}[\ell'']\in \mathbb{C}^{N_{\text{RX},b}\times N_{\text{TX},b}}$, $\forall \ell'' = \{0,1,\ldots,L_{\rm C}-1\}$ represents the coefficients of the wideband analog SI canceller, which is modeled as an $L_{\rm C}$th order Finite Impulse Response (FIR) filter and will be described in the following Sec.~\ref{sec: analog_canceller}. {Recall that the wideband analog canceller utilizes the TX RF chain output $\widetilde{\x}_b[k]$, which contains the transmitter nonlinear impairments $\z_b[k]$, as defined in \eqref{eq: signal_impairments}. Therefore, the analog canceller is capable of suppressing both linear and nonlinear SI components.}

After the downconversion of the received signals at node $b$, the RF chain outputs are added with the digital SI cancellation signal. The resulting signals are then transformed to frequency domain using the FFT operation, as shown in Fig$.$~\ref{fig: transceiver}. Assuming that the digital cancellation signal at the discrete time instant $k$ is given by  $\d_b[k]\in \mathbb{C}^{N_{\text{RX},b}\times 1}$ and using \eqref{eq: ifft_b}, \eqref{eq: signal_DL}, as well as \eqref{eq: received_b}, the frequency-domain received signal vector $\mathpzc{r}_{b,n}\in \mathbb{C}^{N_{\text{RX},b}\times 1}$ at the $n$th subcarrier of the FFT output can be expressed as 
\begin{align}\label{eq: fftOutput_b}
        \nonumber\mathpzc{r}_{b,n} \triangleq & \frac{1}{\sqrt{N_c}}\sum\limits_{k=0}^{N_c-1} \left(\y_b[k] + \d_b[k]\right) e^{-\frac{j2\pi kn}{N_c}}\\   
        \nonumber= & \frac{1}{\sqrt{N_c}}\sum\limits_{\ell=0}^{L_{\rm UL}-1} \H_{\rm UL}[\ell]\sum\limits_{k=0}^{N_c-1}\Bigg(\mathbf{G}_{1,m_2}\Bigg(\frac{1}{\sqrt{N_c}} \sum\limits_{p=0}^{N_c-1} \mathbfcal{V}_{m_2,p}\mathpzc{s}_{m_2,p} e^{\frac{j2\pi p(k-\ell)}{N_c}}\Bigg)+ \z_{m_2}[k-\ell]\Bigg)e^{-\frac{j2\pi kn}{N_c}}\\
        \nonumber& + \frac{1}{\sqrt{N_c}}\Bigg(\sum\limits_{\ell=0}^{L_{\rm SI}-1} \H_{\rm SI}[\ell] + \sum\limits_{\ell=0}^{L_{\rm C}-1} \C_{b}[\ell]\Bigg)\Bigg(\sum\limits_{k=0}^{N_c-1}\mathbf{G}_{1,b}\Bigg(\frac{1}{\sqrt{N_c}} \sum\limits_{p=0}^{N_c-1} \mathbfcal{V}_{b,p}\mathpzc{s}_{b,p} e^{\frac{j2\pi p(k-\ell)}{N_c}}\Bigg) \\
        &+ \sum\limits_{k=0}^{N_c-1} \z_b[k-\ell]\Bigg)e^{-\frac{j2\pi kn}{N_c}} + \frac{1}{\sqrt{N_c}}\sum\limits_{k=0}^{N_c-1} \d_b[k] e^{-\frac{j2\pi kn}{N_c}} + \frac{1}{\sqrt{N_c}}\sum\limits_{k=0}^{N_c-1} \w_b[k] e^{-\frac{j2\pi kn}{N_c}},\\
        \nonumber=&\; \mathbfcal{H}_{{\rm UL},n}(\G_{1,m_2}\mathbfcal{V}_{m_2,n}\mathpzc{s}_{m_2,n} +\mathpzc{z}_{{m_2},n})+ \left(\mathbfcal{H}_{{\rm SI},n}+{\mathbfcal{C}}_{b,n}\right)\left(\G_{1,b}\mathbfcal{V}_{b,n}\mathpzc{s}_{b,n} + \mathpzc{z}_{b,n}\right) + {\mathpzc{d}}_{b,n} + \mathpzc{w}_{b,n},
\end{align}
where
\begin{equation}\label{eq: freq_rep}
    \begin{split}
        \mathbfcal{H}_{{\rm UL},n}&\triangleq\sum\limits_{l=0}^{L_{\rm UL}-1} \H_{\rm UL}[l]e^{-\frac{j2\pi ln}{N_c}},\quad \mathbfcal{H}_{{\rm SI},n}\triangleq\sum\limits_{l=0}^{L_{\rm SI}-1} \H_{\rm SI}[l]e^{-\frac{j2\pi ln}{N_c}},\quad
        \mathbfcal{C}_{b,n}\triangleq\sum\limits_{l=0}^{L_{\rm C}-1} \C_{b}[l]e^{-\frac{j2\pi ln}{N_c}}
    \end{split}
\end{equation}
denote the frequency domain representations of the UL, SI channel, and the wideband analog SI canceller, respectively. Similarly, ${\mathpzc{z}}_{b,n}$, ${\mathpzc{z}}_{m_2,n}$, $\mathpzc{d}_{b,n}$, and $\mathpzc{w}_{b,n}$ are defined as the $n$th subcarrier frequency transform of the TX nonlinear components vector $\z_{b}[k]$ and $\z_{m_2}[k]$, digital cancellation signal vector $\d_b[k]$, and the AWGN vector $\w_{b}[k]$, respectively.

At the FFT output, the $n$th subcarrier symbol vector is linearly processed by the RX combiner $\mathbfcal{U}_{b,n}\in \mathbb{C}^{d_{m_2}\times N_{\text{RX},b}}$ to obtain the estimated symbol vector $\widehat{\mathpzc{s}}_{m_2,n}\in \mathbb{C}^{d_{m_2}\times 1}$, which is derived as
\begin{equation}\label{eq: est_s_m}
    \begin{split}
        \widehat{\mathpzc{s}}_{m_2,n} \triangleq &\; \mathbfcal{U}_{b,n}\big(\mathbfcal{H}_{{\rm UL},n}(\G_{1,m_2}\mathbfcal{V}_{m_2,n}\mathpzc{s}_{m_2,n} +\mathpzc{z}_{{m_2},n}) \\&+ \left(\mathbfcal{H}_{{\rm SI},n}+{\mathbfcal{C}}_{b,n}\right)\left(\G_{1,b}\mathbfcal{V}_{b,n}{\mathpzc{s}}_{b,n} + {\mathpzc{z}}_{b,n}\right) + \mathpzc{d}_{b,n} + \mathpzc{w}_{b,n}\big).
    \end{split}
\end{equation}
\section{Join Digital TX/RX Beamforming and Wideband Analog Cancellation}\label{sec: analog_canceller}
In this section, we present the joint design of digital TX/RX beamforming with wideband analog SI cancellation. We first describe the proposed analog SI canceller for wideband FD MIMO OFDM radios and then present the mathematical formulation for the co-design of the analog cancellation matrix with the digital TX/RX beamformers.

\subsection{Wideband FD MIMO Analog SI Canceller}
Upon signal reception at the FD MIMO node $b$, analog SI cancellation is applied to the signals received at the RX antennas before entering the RF chains, as shown in Fig.~\ref{fig: transceiver}. As previously described in the considered signal model, the wideband analog SI canceller intended for suppressing the multipath SI channel $\H_{\rm SI}[\ell]$ with $\ell=\{0,1,\ldots,L_{\rm SI}-1\}$ is modeled as an $L_{\rm C}$th order FIR filter with the coefficients $\C_b[\ell]$, $\forall \ell=\{0,1,\ldots,L_{\rm C}-1\}$. This filter takes the outputs of the TX RF chains as inputs and routes its outputs to the inputs of the RX RF chains. Note that the special case of $L_{\rm C}=1$ was considered in \cite{alexandropoulos2017joint,islam2019unified,cao2020integrated} for suppressing narrowband SI signals. As will be shown in Sec.~\ref{sec: Simulation} with the performance evaluation results, such narrowband SI cancellers are not capable of suppressing the SI power below the maximum RX input power limit, and therefore, lead to RX RF chains' saturation. To avoid such saturation in wideband FD MIMO systems, full-tap wideband analog SI cancellation was utilized in \cite{antonio2013adaptive,bharadia2014full} that requires $L_{\rm C}\geq L_{\rm SI}$. It is apparent that the hardware complexity of the full-tap cancellers scales with the number of TX/RX antenna elements as well as the multipath channel components.

\begin{figure*}[!t]
\centering
\includegraphics[width=1\textwidth]{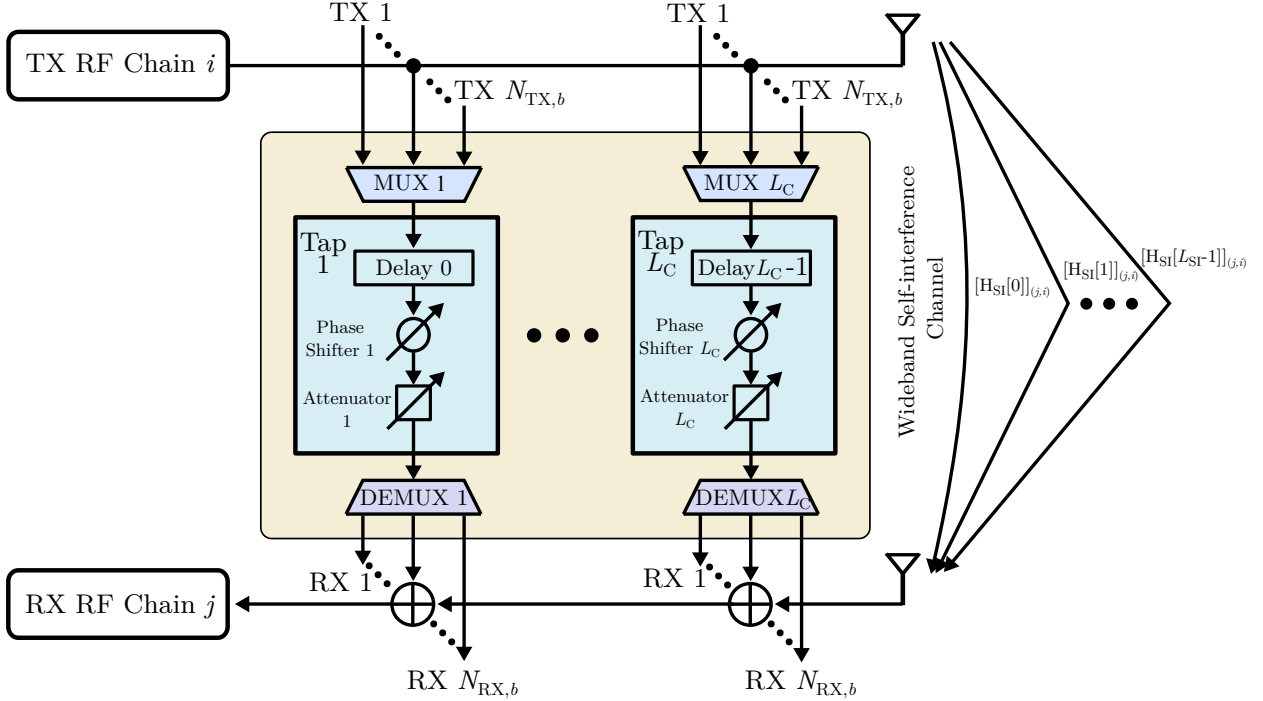}
\caption{The proposed wideband analog SI canceller for the FD MIMO OFDM node $b$ for mitigating the $L_{\rm SI}$ channel paths $[\mathbf{H}_{\rm SI}[\ell]]_{j,i}$ $\forall$ $\ell=\{0,1,\ldots,L_{\rm SI}-1\}$ between the output of the $i$th TX RF chain and the input of the $j$th RX RF chain with $i=\{1,2,\ldots,N_{{\rm TX},b}\}$ and $j=\{1,2,\ldots,N_{{\rm RX},b}\}$.}
\label{fig: canceller}
\end{figure*}

The proposed wideband analog SI canceller for the FD MIMO OFDM node $b$ is depicted in Fig$.$~\ref{fig: canceller}. As illustrated in the figure, it consists of $L_{\rm C}$ analog taps, each including a delay line, a phase shifter, and an attenuator, to suppress the SI signal between the output of any $i$th TX RF chain and the input of any $j$th RX RF chain with $i=\{1,2,\ldots,N_{{\rm TX},b}\}$ and $j=\{1,2,\ldots,N_{{\rm RX},b}\}$. It will be shown in the sequel that the proposed co-design of the analog SI canceller with the digital TX/RX beamformers allows choosing $L_{\rm C}<L_{\rm SI}$, thus reducing the hardware complexity of the canceller compared to the full-tap analog canceller case \cite{antonio2013adaptive,bharadia2014full} in state-of-the-arts. As shown in Fig$.$~\ref{fig: canceller}, outputs of the TX RF chains are routed to the analog canceller via MUltipleXers (MUXs), and the outputs of the canceller are added to selected RX RF chains input using DEMUltipleXers (DEMUXs). The analog canceller settings in Fig$.$~\ref{fig: canceller} are repeated for all the TX and RX RF chains resulting in $N$-tap wideband analog SI canceller. Therefore, for the $\ell$th filter delay, the baseband representation $\C_{b}[\ell]$ of the $N$-tap analog canceller is modeled as
\begin{equation}\label{eq:SI_canc}
    \begin{split}
        \C_{b}[\ell] \triangleq \L_3[\ell] \L_2[\ell] \L_1[\ell],\quad \forall \ell= \{0,1,\ldots,L_{\rm C}-1\},
    \end{split}
\end{equation}
where $\L_1[\ell]\in \mathbb{R}^{\frac{N}{L_{\rm C}} \times N_{{\rm TX},b}}$ and $\L_3[\ell]\in \mathbb{R}^{N_{{\rm RX},b}\times \frac{N}{L_{\rm C}}}$ represent the MUX and DEMUX configurations of the $\ell$th order of the canceller, respectively, and they take the binary values $0$ or $1$. Therefore, it must hold that $\sum\limits_{j=1}^{N_{{\rm TX},b}} [\L_1[\ell]]_{i,j} = 1$ and $\sum\limits_{i=1}^{N_{{\rm RX},b}} [\L_3[\ell]]_{i,j} = 1$,$\forall\, i,j=\{1,2,\dots,\frac{N}{L_{\rm C}}\}$. Here, $\L_2[\ell]\in \mathbb{C}^{\frac{N}{L_{\rm C}} \times \frac{N}{L_{\rm C}}}$ is a diagonal matrix whose complex entries represent the attenuation and phase shift of the $\ell$th sample delayed canceller taps. {For example, we consider a $4\times4$ wideband MIMO system with $\frac{N}{L_{\rm C}} = 12$ available analog cancellation taps at the $\ell$th filter delay. One possible MUX and DEMUX configuration of the $\ell$th order of the canceller can be written as $\L_1[\ell] = [\I_4\,\,\I_4 \,\,\I_4]^{\rm T}$ and $\L_3[\ell] = [\I_4\,\,\I_4 \,\,\I_4]$, respectively. Evidently, the above constraints are satisfied for the MUX and DEMUX configurations.
}

{
Compared to the narrowband \cite{alexandropoulos2017joint,islam2019unified,cao2020integrated} and full-tap wideband cancellers \cite{antonio2013adaptive,bharadia2014full}, our proposed wideband analog SI canceller reduces the complexity in two ways: firstly, the analog canceller $\C_b[\ell]$ has filter order $L_{\rm C}\leq L_{\rm SI}$, and secondly, the canceller is capable of selecting the minimum number of TX/RX antenna pairs, whose SI impact is to be suppressed to avoid the RX RF chains' saturation. Therefore, the hardware complexity of the proposed wideband analog SI canceller does not scale with the number of antennas nor with the number of multipath components. Assuming $N$ as the total number of taps of the canceller $\C_b[\ell],\, \forall \ell$, it holds that $N\leq N_{{\rm RX},b} N_{{\rm TX},b} L_{\rm C}\leq N_{{\rm RX},b} N_{{\rm TX},b} L_{\rm SI}$.}

\subsection{Digital TX/RX Beamforming and Wideband Analog SI Cancellation}\label{subsec: analog_canellation}
Suppose that the UL, DL, and SI wireless channels in the considered system of Fig$.$~\ref{fig: transceiver} are estimated using pilot signals as $\widehat{\H}_{\rm DL}[\ell]$, $\widehat{\H}_{\rm UL}[\ell]$, and $\widehat{\H}_{\rm SI}[\ell]$, $\forall \ell$, respectively. Using these estimations and the representation for the analog canceller as well as digital TX/RX beamformers, estimates for the $n$th subcarrier achievable UL and DL rates can be respectively calculated as
\begin{align}\label{eq:rates_eq}
    % \begin{split}
        \nonumber\widehat{\mathcal{R}}_{\text{UL},n}&\triangleq\log_2\left(\det\left(\mathbf{I}_{d_{m_2}}+ \left\|\mathbfcal{U}_{b,n}\widehat{\mathbfcal{H}}_{{\rm UL},n}\G_{1,m_2}\mathbfcal{V}_{m_2,n}\right\|^2{\widehat{\mathbfcal{Q}}_{b,n}}^{-1}\right)\right),\\
        \widehat{\mathcal{R}}_{\text{DL},n}&\triangleq\log_2\left(\det\left(\mathbf{I}_{d_b}+ \left\|\mathbfcal{U}_{m_1,n}\widehat{\mathbfcal{H}}_{{\rm DL},n}\G_{1,b}\mathbfcal{V}_{b,n}\right\|^2{\widehat{\mathbfcal{Q}}_{m_1,n}}^{-1}\right)\right),
    % \end{split}
\end{align}
where $\widehat{\mathbfcal{Q}}_{b,n}$ and $\widehat{\mathbfcal{Q}}_{m_1,n}$ denote the estimated Interference-plus-Noise (IpN) covariances matrices at multi-antenna nodes $b$ and $m_1$, respectively, which can be computed as
\begin{equation}\label{eq:covar_eq}
    \begin{split}
        &\widehat{\mathbfcal{Q}}_{b,n} \triangleq \left\|\mathbfcal{U}_{b,n}\left({\widetilde{\mathbfcal{H}}}_{{\rm SI},n}\left(\G_{1,b}\mathbfcal{V}_{b,n}{\mathpzc{s}}_{b,n} + \mathpzc{z}_{b,n}\right) + \widehat{\mathbfcal{H}}_{{\rm UL},n} \mathpzc{z}_{m_2,n} + {\mathpzc{d}}_{n} \right)\right\|^2 + \sigma_b^2\|\mathbfcal{U}_{b,n}\|^2,\\
        &\widehat{\mathbfcal{Q}}_{m_1,n} \triangleq \left\|\mathbfcal{U}_{m_1,n}\left(\widehat{\mathbfcal{H}}_{{\rm DL},n} \mathpzc{z}_{b,n}\right)\right\|^2 + \sigma_{m_1}^2\|\mathbfcal{U}_{m_1,n}\|^2,
    \end{split}
\end{equation}
where ${\widetilde{\mathbfcal{H}}}_{{\rm SI},n}\triangleq\left(\widehat{\mathbfcal{H}}_{{\rm SI},n}+{\mathbfcal{C}}_{b,n}\right)$. The latter expressions have been obtained from \eqref{eq: est_s_m} and \eqref{eq: est_s_b} assuming estimation of the TX impairments at RXs.

% and that ${\mathpzc{z}}_{b,n} =\0_{N_{{\rm RX},m_1}\times1}$ (i$.$e$.$, yet no digital SI cancellation).\par

Extending the design approach of \cite{islam2019unified}, we focus on the estimated achievable FD rate of $n$th subcarrier defined as the sum of $\widehat{\mathcal{R}}_{\text{UL},n}$ and $\widehat{\mathcal{R}}_{\text{DL},n}$, and formulate the following general optimization problem for the joint design of the $N$-tap wideband analog SI canceller and the digital TX/RX beamformers: 
\begin{align}\label{eq:optimization_eq}
    %\begin{split}
        \nonumber\mathcal{OP}1:\underset{\substack{\C_{b}[\ell], \mathbfcal{V}_{b,n},  \mathbfcal{U}_{b,n},\\\mathbfcal{V}_{m_2,n},\mathbfcal{U}_{m_1,n}, {\mathpzc{d}}_{b,n}}}{\text{max}} &\widehat{\mathcal{R}}_{\text{UL},n}+\widehat{\mathcal{R}}_{\text{DL},n}&\\
        \text{\text{s}.\text{t}}\quad\quad & \text{constraints}\, \text{on}\, \C_{b}[\ell], \forall \ell\, \text{structure},& \text{(C1)}\nonumber\\
        &\mathbb{E}\{\|\G_{1,b}\mathbfcal{V}_{b,n}{\mathpzc{s}}_{b,n} +\mathpzc{z}_{b,n}\|^2\}\leq {\rm P}_b,& \text{(C2)}\nonumber\\
        &\mathbb{E}\{\|\G_{1,m_2}\mathbfcal{V}_{m_2,n}{\mathpzc{s}}_{m_2,n} +\mathpzc{z}_{m_2,n} \|^2\}\leq {\rm P}_{m_2},& \text{(C3)}\nonumber\\
        &\mathbfcal{V}_{b,n},\mathbfcal{U}_{b,n},\mathbfcal{V}_{m_2,n},\mathbfcal{U}_{m_1,n}:\,\text{unit}\,\,\text{norm}\,\,\text{columns},& \text{(C4)}\nonumber\\
        & \left\|{\widetilde{\mathbfcal{H}}}_{{\rm SI},n}\left(\G_{1,b}\mathbfcal{V}_{b,n}{\mathpzc{s}}_{b,n} + \mathpzc{z}_{b,n}\right)\right\|^2<\lambda_{b}\I_{N_{{\rm RX},b}},& \text{(C5)}\nonumber
    %\end{split}
\end{align}
where $\text{(C1)}$ represents the analog SI canceller settings as in \eqref{eq:SI_canc}, constraints $\text{(C2)}, \text{(C3)}$ relate to the average transmit power at node $b$ and $m_2$, respectively, $\text{(C4)}$ enforces the unit norm condition of the considered beamformers, and $\text{(C5)}$ imposes the threshold residual power level, $\lambda_{b}$ at the RX antenna input to avoid RF chain saturation at node $b$. The threshold power is limited by the ADC dynamic range.\par

\begin{algorithm}[!t]
    \caption{Digital TX/RX Beamforming Maximizing DL Rate}
    \label{alg:DL_alg}
    \begin{algorithmic}[1]
        \renewcommand{\algorithmicrequire}{\textbf{Input:}}
      \renewcommand{\algorithmicensure}{\textbf{Output:}}
        \Require $\C_b[\ell], \forall \ell$, ${\rm P}_b$, $\mu_1$, $\nu_1$, $\mu_2$, $\nu_3$, $\sigma_{m_1}^2$, and $n$.
        \Ensure $\mathbfcal{V}_{b,n}$, $\G_{1,b}$, and $\mathbfcal{U}_{m_1,n}$.
        \State Obtain wireless channel estimates $\widehat{\H}_{\rm DL}[\ell]$ and $\widehat{\H}_{\rm SI}[\ell]$ using pilot signals.
        \State Get $\widehat{\mathbfcal{H}}_{{\rm DL},n}$, $\widehat{\mathbfcal{H}}_{{\rm SI},n}$ and $\mathbfcal{C}_{b,n}$ using \eqref{eq: freq_rep}.
        \State Obtain $\mathbfcal{D}_b$ including the $N_{{\rm TX},b}$ right-singular vectors of ${\widetilde{\mathbfcal{H}}}_{{\rm SI},n}=\left(\widehat{\mathbfcal{H}}_{{\rm SI},n}+{\mathbfcal{C}}_{b,n}\right)$ corresponding to the singular values in descending order.
        \State Set  $\alpha_{max}=\text{min}\{N_{{\rm RX},m_1},N_{{\rm TX},b}\}$.
        \For{$\alpha=\alpha_{max},\alpha_{max}-1,\dots,2$}
            \State Set $\mathbfcal{E}_b = [\mathbfcal{D}_b]_{(:,N_{{\rm TX},b}-\alpha+1:N_{{\rm TX},b})}$.
            \State Set $\mathbfcal{F}_b$ as the right singular vectors of effective DL channel $\widehat{\mathbfcal{H}}_{{\rm DL},n}\mathbfcal{E}_b$.
            \State Set $\mathbfcal{V}_{b,n}=\mathbfcal{E}_b\mathbfcal{F}_b$.
            \State Set $[\G_{1,b}]_{(i,i)}=\sqrt{{\rm P}_b/N_{{\rm TX},b}}, \forall i=1,2,\dots,N_{{\rm TX},b}$.
            \State Obtain $\mathbf{G}_{j,b}$ for $j=2,3,\ldots,6$ using \eqref{eq: small_gs}.
            \If {$\Big(\Big\|\left[{\widetilde{\mathbfcal{H}}}_{{\rm SI},n}\G_{1,b} \mathbfcal{V}_{b,n}\right]_{(i,:)}\Big\|^2 + \left|\left[{\widetilde{\mathbfcal{H}}}_{{\rm SI},n}\mathpzc{z}_{b,n}\right]_{i}\right|^2 \Big)<\lambda_{b}$}
                \State Stop the loop.
            \EndIf
        \EndFor
        \State Set $\mathbfcal{U}_{m_1,n}$ as row-wise placement of the $d_b$ left singular vectors of effective DL channel $\widehat{\mathbfcal{H}}_{{\rm DL},n}\mathbfcal{E}_b$ corresponding to the singular values in descending order.
        \If{$\Big(\Big\|\left[{\widetilde{\mathbfcal{H}}}_{{\rm SI},n}\G_{1,b} \mathbfcal{V}_{b,n}\right]_{(i,:)}\Big\|^2 + \left|\left[{\widetilde{\mathbfcal{H}}}_{{\rm SI},n}\mathpzc{z}_{b,n}\right]_{i}\right|^2 \Big)<\lambda_{b}$}
            \State Output $\mathbfcal{V}_{b,n}$, $\G_{1,b}$, $\mathbfcal{U}_{m_1,n}$, and stop the algorithm.
        \Else
            \State Numbers of taps, $N$ in analog canceller $\C_b[\ell]$ is not capable of preventing RF saturation.
        \EndIf
    \end{algorithmic}
\end{algorithm}

The optimization problem in $\mathcal{OP}1$ is a nonconvex problem with nonconvex constraints. We propose to solve $\mathcal{OP}1$ in a decoupled way, where two subproblems are formulated and solved subject to the constraints. First, we solve for  $\C_{b}[\ell],\mathbfcal{V}_{b,n}$, and $\mathbfcal{U}_{m_1,n}$ that maximize the instantaneous DL rate subject to the relevant constraints for these unknown variables and the constraint to ensure that RF saturation is satisfied at all node $b$ RXs. More specifically, we formulate the following optimization subproblem for the design
\begin{equation}\label{eq:DL_optimization}
    \begin{split}
        \nonumber\mathcal{OP}2:\underset{\substack{\C_{b}[\ell],\mathbfcal{V}_{b,n},\mathbfcal{U}_{m_1,n}}}{\text{max}} & \widehat{\mathcal{R}}_{\text{DL},n}\\
        \text{\text{s}.\text{t}}\quad\quad & \text{(C1)},\text{(C2)},\text{(C4)},\text{(C5)}\nonumber
    \end{split}
\end{equation}
We solve the problem $\mathcal{OP}2$ adopting an alternating optimization approach. Specifically, supposing that the available number of analog canceller taps $N$ and a realization of $\C_{b}[\ell],\,\forall \ell$ satisfying the constraint $\text{(C1)}$ are given, we seek for $\mathbfcal{V}_{b,n}$ and $\mathbfcal{U}_{m_1,n}$ maximizing the DL rate while meeting the constraints $\text{(C2)},\text{(C4)}$ and the constraint $\text{(C5)}$ for the residual SI after analog cancellation. The latter procedure is repeated for all allowable realizations of $\C_{b}[\ell],\,\forall \ell$ for the given $N$ in order to find the best variables $\C_{b}[\ell],\mathbfcal{V}_{b,n},\mathbfcal{U}_{m_1,n}$ solving $\mathcal{OP}2$.
For a given number of analog cancellation taps $N$, we define the analog cancellation filter order $L_{\rm C}=\lceil \frac{N}{N_{{\rm TX},b} N_{{\rm RX},b}} \rceil$. Now we formulate the analog canceller $\C_{b}[\ell]$ as an orderly column-by-column placement of the reciprocal of wideband SI channel $\widehat{\H}_{\rm SI}[\ell]$ elements.
Based on this analog canceller $\C_{b}[\ell],\,\forall \ell$, we follow the procedures summarized in Algorithm~\ref{alg:DL_alg}, solving for the beamformers maximizing the DL rate.

Using $\C_{b}[\ell],\,\forall \ell$, $\mathbfcal{V}_{b,n}$, and $\mathbfcal{U}_{m_1,n}$ from the solution of $\mathcal{OP}2$, we transmit $T$ precoded training symbols to obtain the digital cancellation signal ${\mathpzc{d}}_{b,n}$ following the procedures described in Sec.~\ref{sec: digital}. Using the pilot estimated UL channel, DL beamformers, analog canceller, and digital cancellation signal, we propose to maximize the instantaneous UL rate for the solution of $\mathcal{OP}2$. In particular, we formulate the following optimization subproblem for the digital beamformers to maximize the UL rate:

\begin{equation}\label{eq:UL_optimization}
    \begin{split}
        \nonumber\mathcal{OP}3:\underset{\substack{\mathbfcal{V}_{m_2,n},\mathbfcal{U}_{b,n}}}{\text{max}} & \widehat{\mathcal{R}}_{\text{UL},n}\\
        \text{\text{s}.\text{t}}\quad\quad & \text{(C3)},\text{(C4)}\nonumber
    \end{split}
\end{equation}
The procedures to solve the optimization problem $\mathcal{OP}3$ is provided in Algorithm~\ref{alg:UL_alg}.

\begin{algorithm}[!t]
    \caption{Digital TX/RX Beamforming Maximizing UL Rate}
    \label{alg:UL_alg}
    \begin{algorithmic}[1]
        \renewcommand{\algorithmicrequire}{\textbf{Input:}}
      \renewcommand{\algorithmicensure}{\textbf{Output:}}
        \Require $\mathbfcal{V}_{b,n}$, ${\widetilde{\mathbfcal{H}}}_{{\rm SI},n}$,${\mathpzc{d}}_{b,n}$, ${\rm P}_{m_2}$, $\sigma_b^2$, and $n$.
        \Ensure $\mathbfcal{V}_{m_2,n}$, $\G_{1,m_2}$, and $\mathbfcal{U}_{b,n}$.
        \State Obtain UL channel estimates $\widehat{\H}_{\rm UL}[\ell]$ using pilot signals, and its $n$-th subcarrier frequency domain presentation $\widehat{\mathbfcal{H}}_{{\rm UL},n}$ using \eqref{eq: freq_rep}.
        \State Obtain $\mathbfcal{D}_{m_2}$ including the $N_{{\rm TX},m_2}$ right-singular vectors of $\widehat{\mathbfcal{H}}_{{\rm UL},n}$ corresponding to the singular values in descending order.
        \State Set $\mathbfcal{V}_{m_2,n} = [\mathbfcal{D}_{m_2}]_{(:,1:d_{m_2})}$.
        \State Set $[\G_{1,m_2}]_{(i,i)}=\sqrt{{\rm P}_{m_2}/N_{{\rm TX},m_2}}, \forall i=1,2,\dots,N_{{\rm TX},m_2}$.
        \State Set $\boldsymbol{\Sigma}_b = {\widetilde{\mathbfcal{H}}}_{{\rm SI},n}\left(\G_{1,b} \mathbfcal{V}_{b,n}\mathbfcal{V}_{b,n}^{\rm H}\G_{1,b}^{\rm H} + \mathpzc{z}_{b,n}\mathpzc{z}_{b,n}^{\rm H}\right)\widetilde{\mathbfcal{H}}_{{\rm SI},n}^{\rm H} + \widehat{\mathbfcal{H}}_{{\rm UL},n} \mathpzc{z}_{m_2,n}\mathpzc{z}_{m_2,n}^{\rm H}\widehat{\mathbfcal{H}}_{{\rm UL},n}^{\rm H} + \mathpzc{d}_{n}{\mathpzc{d}_{n}}^{\rm H} + \sigma_b^2\I_{N_{{\rm RX},b}}$.
        \State Set $\mathbfcal{A}_b =\widehat{\mathbfcal{H}}_{{\rm UL},n}\G_{1,m_2}\mathbfcal{V}_{m_2,n}\mathbfcal{V}_{m_2,n}^{\rm H}\G_{1,m_2}^{\rm H}\widehat{\mathbfcal{H}}_{{\rm UL},n}^{\rm H}\boldsymbol{\Sigma}_b^{-1}$.
        \State Set $\mathbfcal{U}_{b,n}$ as row-wise placement of the $d_{m_2}$ eigenvectors of $\mathbfcal{A}_b$ corresponding to $d_{m_2}$ largest eigenvalues.
    \end{algorithmic}
\end{algorithm}

\section{TSVD-based Adaptive MIMO Digital SI Cancellation}\label{sec: digital}
To suppress the residual SI signal after analog cancellation, digital SI mitigation techniques are employed at the RX baseband of the FD node. The existing digital cancellation approaches utilize the SI signal modeling to estimate the linear and/or nonlinear SI components resulting from the FD TX RF chains to reconstruct the residual SI signal \cite{korpi2014widely,bharadia2014full,anttila2014modeling,korpi2017nonlinear}. However, for the considered wideband FD MIMO systems, the number of signal estimation parameters of those models increases with   TX/RX  RF chains and multipath SI channel components, which requires a large number of training signals. In this section, we propose a novel TSVD-based adaptive MIMO digital SI cancellation that reduces the computational complexity while successfully suppressing the residual SI signal.

At the node $b$ RX, the baseband received signal $\y_{b}[k]$ contains the desired signal from the UL TX node $m_2$, residual SI signal from node $b$ after analog cancellation, and the AWGN vector. From \eqref{eq: signal_DL}, \eqref{eq:augment_signal}, and \eqref{eq: received_b}, the residual SI signal with AWGN vector can be written as
\begin{equation}\label{eq: res_sig_dig}
    \begin{split}
        \y_{\rm res} [k] &\triangleq \sum\limits_{\ell = 0}^{L_{\rm SI}-1} \H_{\rm res}[\ell] \boldsymbol{\psi}_{b}[k-\ell] + \w_b[k]\\
        &= \widetilde{\H}_{\rm res} \widetilde{\boldsymbol{\psi}}_{b}[k] + \w_b[k],
    \end{split}
\end{equation}
where $\H_{\rm res}[\ell] = (\H_{\rm SI}[\ell] + \C_{b}[\ell])\G_b$ is the residual SI channel and $\boldsymbol{\psi}_{b}[k]$ is the vertically arranged TX precoded signal vector containing combination of $\x_{b}[k]$, $\x^*_{b}[k]$, and their third-order components as defined in \eqref{eq:augment_signal}. Here the augmented residual SI channel $\widetilde{\H}_{\rm res}\in \mathbb{C}^{N_{{\rm RX},b}\times 6N_{{\rm TX},b}L_{\rm SI}}$ and the precoded signal vector $\widetilde{\boldsymbol{\psi}}_{b}[k]\in \mathbb{C}^{6N_{{\rm TX},b}L_{\rm SI}\times 1}$ are defined as
\begin{equation}\label{eq: aug_sig_dig}
    \begin{split}
        \widetilde{\H}_{\rm res} &\triangleq [\H_{\rm res}[0]\,\, \H_{\rm res}[1]\,\,\ldots\,\, \H_{\rm res}[L_{\rm SI}-1]]\\
        \widetilde{\boldsymbol{\psi}}_{b}[k] &\triangleq {\rm col}\{\boldsymbol{\psi}_{b}[k],\boldsymbol{\psi}_{b}[k-1],\ldots,\boldsymbol{\psi}_{b}[k-L_{\rm SI}+1]\}
    \end{split}
\end{equation}

The objective of the digital SI cancellation is to estimate the residual SI channel parameters of $\widetilde{\H}_{\rm res}$ using training samples during no UL communication and utilize these estimated parameters to reconstruct the reciprocal of the residual SI signal during FD operation. The number of required training samples to estimate the precise residual SI channel parameters ($N_{{\rm RX},b}\times 6N_{{\rm TX},b}L_{\rm SI}$) increases with the number of FD MIMO node $b$ antennas and the multipath SI components. To reduce the requirement of a large number of training signals and computational resources, we propose a novel TSVD-based adaptive digital SI cancellation approach for wideband FD MIMO systems.

\begin{algorithm}[!t]
    \caption{Adaptive Digital SI cancellation using TSVD regularization}
    \label{alg:adap_dig_can}
    \begin{algorithmic}[1]
        \renewcommand{\algorithmicrequire}{\textbf{Input:}}
        \renewcommand{\algorithmicensure}{\textbf{Output:}}
        \Require $\X_b \in \mathbb{C}^{N_{{\rm TX},b}\times T}, \Y_{\rm res}\in \mathbb{C}^{N_{{\rm RX},b}\times T}, \sigma_b^2$.
        \Ensure $\widehat{\widetilde{\H}}_{\rm res}\in \mathbb{C}^{N_{{\rm RX},b}\times 6N_{{\rm TX},b}L_{\rm SI}}$.
        \State Obtain $\widetilde{\boldsymbol{\Psi}}_{b}$ using \eqref{eq: aug_sig_dig}.
        \State Set $L_p = 6N_{{\rm TX},b}L_{\rm SI}$.
        \State Obtain SVD of $\widetilde{\boldsymbol{\Psi}}_{b} = \mathbf{U}\boldsymbol{\Sigma}\mathbf{V}^{\rm H}$, where $\mathbf{U}\in \mathbb{C}^{L_p\times L_p}$, $\mathbf{V}\in \mathbb{C}^{T\times T}$, $\boldsymbol{\Sigma}=\text{diag}\{\sigma_1, \sigma_2,\dots,\sigma_{L_p}\} \in \mathbb{C}^{L_p\times T}$, and $\sigma_1\geq \sigma_2\geq\dots\geq\sigma_{L_p}\geq 0$.
        \For{$p=1,2,\ldots,L_p$}
            \State Set $\boldsymbol{\Theta} = \sum\limits_{i=1}^{p} \frac{\Y_{\rm res} [\V]_{(:,i)}[\U]_{(:,i)}^{\rm H}}{\sigma_i} $.
            \If{$\|\Y_{\rm res}-\boldsymbol{\Theta}\widetilde{\boldsymbol{\Psi}}_{b}\|^2\leq \sigma_b^2\I_{{\rm RX},b}$}
                \State Set $\widehat{\widetilde{\H}}_{\rm res}=\boldsymbol{\Theta}$.
                \State \textbf{Stop the algorithm.}
            \EndIf
        \EndFor
        \State Set $\widehat{\widetilde{\H}}_{\rm res}=\boldsymbol{\Theta}$.
    \end{algorithmic}
\end{algorithm}

\subsection{Proposed Digital SI Cancellation}
To design the digital SI cancellation, we utilize the precoded signal vector at the baseband of node $b$ TXs. Supposing $T$ training samples, the precoded signal matrix $\X_b \in \mathbb{C}^{N_{{\rm TX},b}\times T}$ is used to form the augmented signal matrix $\widetilde{\boldsymbol{\Psi}}_b\in \mathbb{C}^{6N_{{\rm TX},b}L_{\rm SI}\times T}$ using \eqref{eq: aug_sig_dig}. Therefore, based on \eqref{eq: res_sig_dig}, the residual SI signal matrix $\Y_{\rm res} \in \mathbb{C}^{N_{{\rm RX},b}\times T}$ with AWGN at the RX baseband of node $b$ is expressed as
\begin{equation}\label{eq: res_mat_dig}
    \begin{split}
        \Y_{\rm res} = \widetilde{\H}_{\rm res} \widetilde{\boldsymbol{\Psi}}_{b} + \W_b
    \end{split}
\end{equation}
Making use of the SI samples in \eqref{eq: res_mat_dig} and the notation $\boldsymbol{\Theta}\in\mathbb{C}^{N_{{\rm RX},b} \times 6N_{{\rm TX},b}L_{\rm SI}}$, the Least Squares (LS) estimation for $\widetilde{\H}_{\rm res}$ minimizing the power of the error matrix can be expressed as
\begin{equation}\label{eq:arg_min_res}
        \widehat{\widetilde{\H}}_{\rm res}\triangleq \argmin_{\boldsymbol{\Theta}}\|\Y_{\rm res}- \boldsymbol{\Theta}\widetilde{\boldsymbol{\Psi}}_b\|^2.
\end{equation}  

The LS problem has a closed form solution as $\widehat{\widetilde{\H}}_{\rm res} = \Y_{\rm res}\widetilde{\boldsymbol{\Psi}}_b^{\rm H}\left(\widetilde{\boldsymbol{\Psi}}_b\widetilde{\boldsymbol{\Psi}}_b^{\rm H}\right)^{-1}$, assuming full row rank in $\widetilde{\boldsymbol{\Psi}}_b$. However, $\widetilde{\boldsymbol{\Psi}}_b$'s rows are high order polynomials of linear and conjugate SI samples, as well as their interaction products, and therefore correlated. In addition, the LS solution includes the effect of strong SI components along with less significant terms and is dominated by estimation error. To tackle the estimation error and reduce the parameter estimation model complexity, we propose the following TSVD-based digital SI cancellation method.

The main idea of the proposed adaptive digital SI cancellation method is to include the SI terms that have a strong effect on the residual SI signal while omitting the insignificant SI terms. We apply the TSVD regularization method to the estimation problem. First, we perform the SVD of the design matrix $\widetilde{\boldsymbol{\Psi}}_b$ to obtain the singular values in an ascending manner. Now, we loop through the singular values of $\widetilde{\boldsymbol{\Psi}}_b$ to find its best rank-$p$ approximant and the effective residual SI channel that can suppress the residual SI signal below the noise floor $\sigma_b^2$. The TSVD-based adaptive digital cancellation approach is described in Algorithm~\ref{alg:adap_dig_can}.

The proposed cancellation approach is adaptive because the values of $p$ vary as the SI signal power changes. For higher SI power, the algorithm estimates more parameters resulting in a large $p$ value. However, for low transmit power, the proposed approach omits the insignificant SI terms with small $p$ values and reduces the estimation error. Using the residual channel estimate $\widehat{\widetilde{\H}}_{\rm res}$, the digital cancellation signal $\d_b[k]$ can be formed as
\begin{equation}\label{eq:digital}
    \begin{split}
        \d_b[k]\triangleq -\widehat{\widetilde{\H}}_{\rm res}\widetilde{\boldsymbol{\psi}}_{b}[k],
    \end{split}
\end{equation}
which is applied to the received signals after the ADCs.\par
\section{Simulation Results and Discussion}\label{sec: Simulation}
In this section, we present the performance evaluation of the proposed wideband FD MIMO scheme incorporating the impairments of the transceiver chains depicted in Fig.~\ref{fig: transceiver}. We also provide a performance comparison with the state-of-the-art A/D SI cancellation schemes relevant to the considered FD MIMO wideband communication system.  
\begin{table}[tbp]
    \caption{Simulation Parameters}
    \label{tab: sim_param}
    \centering
    % \scriptsize
    \begin{tabular}{|c|c|}
         \hline
         \textbf{Parameter} & \textbf{Value} \\
         \hline
         Signal Bandwidth & $20$ MHz \\
         Subcarrier Spacing & $312.5$ KHz\\
         Sampling Time & $50$ns\\
         Constellation & $16$-QAM\\
         No. of Subcarriers &  $64$\\
         No. of Data Subcarriers &  $52$\\
         Cyclic Prefix Length & $16$\\
         \hline
    \end{tabular}
        \!\!\!
    \begin{tabular}{|c|c|}
         \hline
         \textbf{Parameter} & \textbf{Value} \\
         \hline
         Transmit Power & $20-40$ dBm\\
         Node $b$ Noise Floor & $-100$ dBm\\
         Node $m_1$ Noise Floor & $-90$ dBm\\
         PA IIP3 & $15$ dBm\\
         TX IRR & $30$ dB\\
        No. of ADC Bits & $14$\\
         PAPR & $10$ dB\\
         \hline
    \end{tabular}
\end{table}
\squeezeup
\squeezeup
\subsection{Simulation Parameters}\label{subsec:sim_param}
We perform an extensive waveform simulation following the wideband FD MIMO architecture illustrated in Fig$.$~\ref{fig: transceiver}, where the practical transceiver components are incorporated using baseband equivalent models. We consider a $4\times 4$ FD MIMO BS node $b$ (i.e. $N_{{\rm TX},b}=N_{{\rm RX},b} =4$), and two different cases for the number of antennas at HD nodes $m_1$ and $m_2$: the single-antenna case (i.e. $N_{{\rm TX},m_1}=N_{{\rm RX},m_2} =1$) and the multi-antenna with $N_{{\rm RX},m_1}=N_{{\rm TX},m_2} =4$. The generated wideband waveforms are OFDM signals with a  BandWidth (BW) of $20$ MHz. Both DL and UL channels are assumed to be wideband block Rayleigh fading channels with $L_{\rm DL} = L_{\rm UL} = 4$ multipath components and an average pathloss of $100$ dB. {The SI channel at the FD MIMO node $b$ is simulated as a wideband Rician fading channel with $L_{\rm SI} = 4$ multipath components, where the direct SI and the three considered reflected paths have $0,50,100$, and $150$ns delay as well as $40, 50, 60$, and $70$ dB pathlosses, respectively \cite{duarte2012experiment}. For the considered waveform, the $50$ns delay between consecutive SI paths constitutes to one sample delay of the OFDM signal.} The additional parameters of the waveforms, along with other system-level parameters, are shown in Table~\ref{tab: sim_param}. To incorporate TX impairments at the FD MIMO node $b$, the TX IRR is considered $30$ dB, which implies that the image signal power is $30$ dB lower than the linear SI signal \cite{korpi2014widely}. The quasi memoryless nonlinear PAs are assumed to have an IIP3 value of $15$ dBm \cite{ad9361}. Each ADC at the RX RF chains of node $b$ is considered to have a $14$-bit resolution with an effective dynamic range of $60$ dB for a Peak-to-Average-Power-Ratio (PAPR) of $10$ dB \cite{AD3241}. Therefore, the residual SI power after analog cancellation at each RX RF chain of node $b$ has to be below $-40$ dBm to avoid RF saturation. {Furthermore, to account for practical mismatch, the wideband analog canceller taps are considered non-ideal with steps of $0.02$ dB for attenuation and $0.13^\circ$ for phase, as in \cite{alexandropoulos2017joint}. Thus, for each tap in our simulations, the phase setting has a random phase error uniformly distributed between $-0.065^\circ$ and $0.065^\circ$. Considering a waveform with $312.5$KHz subcarrier spacing, the phase error translates to a maximum of $0.578$ns time delay error for each tap.}
We have used $1000$ Monte Carlo simulation runs ($1000$ independent set of channels) to calculate the performance of all considered designs. In each run, we have performed pilot-assisted estimations of all involved wireless channels and considered $500$ OFDM symbols transmitted from both nodes $b$ and $m_2$ to emulate one radio frame for packet-based FD MIMO communication system.
\begin{figure}[!tbp]
\centering
\includegraphics[width=0.7\textwidth]{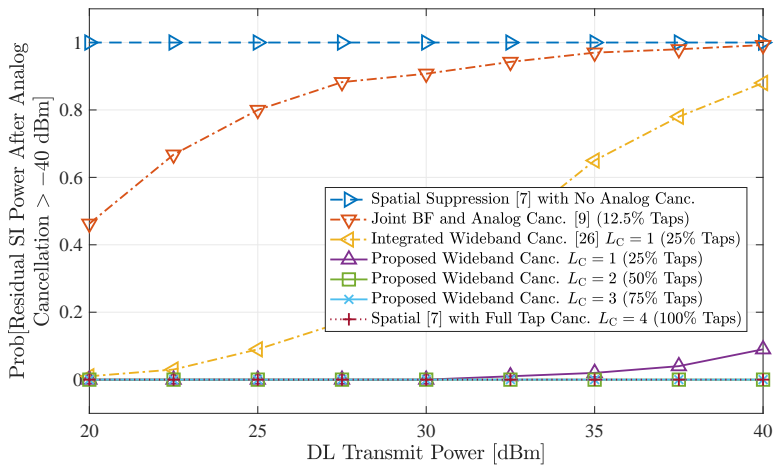}
\caption{Probability of the residual SI power after analog cancellation above RF saturation level $-40$ dBm with respect to DL transmit power in dBm and $N_{{\rm RX},m_1} = N_{{\rm TX},m_2}=1$.}
\label{fig: probSat_single}
\end{figure}
\subsection{Analog SI Mitigation Capability}
First, we evaluate the SI suppression capability of the proposed wideband FD MIMO analog SI canceller and compare it with the state-of-the-art SI cancellation approaches. As discussed in Sec.~\ref{sec: analog_canceller}, the performance indicator of the analog canceller is its ability to avoid RX RF chain saturation at the FD node. To reduce analog hardware complexity, most of the existing methods only deal with direct SI signals in analog cancellation and suppress the multipath components in the digital domain, assuming the insignificant impact of reflected SI components on RF chain saturation.
However, for wideband FD MIMO systems, the multipath SI components can result in residual SI power above the RF saturation level for high transmit powers \cite{wu2014power,askar2019polarimetric,chen2018self}. For example, we have considered a wideband SI channel with reflected paths having pathlosses as high as $70$ dB, which results in an SI power of $-30$ dBm at the RX input of FD MIMO node $b$ for transmit power of $40$ dBm. This residual SI power is above the RF chain saturation level of $-40$ dBm of the considered FD MIMO system, and therefore needs to be suppressed in the analog domain.

In Fig.~\ref{fig: probSat_single}, we illustrate the probability of the residual SI signal power above the RF saturation level of $-40$ dBm after proposed wideband analog cancellation with varying number of taps ($L_{\rm C} = {1,2,3}$) as a function of DL transmit power ($20-40$ dBm) for the considered $4\times 4$ wideband FD MIMO communication system with single-antenna HD users (i.e. $N_{{\rm RX},m_1} = N_{{\rm TX},m_2}=1$). For comparison, we also plot the analog SI cancellation performance of the ``Spatial Suppression" \cite{riihonen2011mitigation}, ``Joint BF and Analog Cancellation" \cite{alexandropoulos2017joint}, and ``Integrated Wideband Cancellation" \cite{cao2020integrated} approaches. It is evident that the ``Spatial Suppression" and ``Joint BF and Analog Cancellation" approaches are unable to achieve enough analog SI suppression to avoid RF saturation, as they only cancel parts of the direct SI components rendering high residual SI power. The ``Integrated Wideband Cancellation" with $L_{\rm C} = 1$ is capable of achieving sufficient analog SI suppression for low transmit power of $20$ dBm; however, for high transmit power cases, the FD MIMO node $b$ RX RF chains go into saturation as the effect of multipath SI components becomes significant. In contrast, the proposed wideband analog cancellation approach with $25\%$ taps ($L_{\rm C} = 1$) can provide sufficient RF suppression for transmit power up to $30$ dBm. After a $50\%$ reduction of taps compared to the ``Spatial Suppression" with full-tap analog canceller, the proposed wideband cancellation with $L_{\rm C} = 2$ is capable of avoiding RF saturation for all DL transmit power values.

\begin{figure}[!tbp]
\centering
\includegraphics[width=0.7\textwidth]{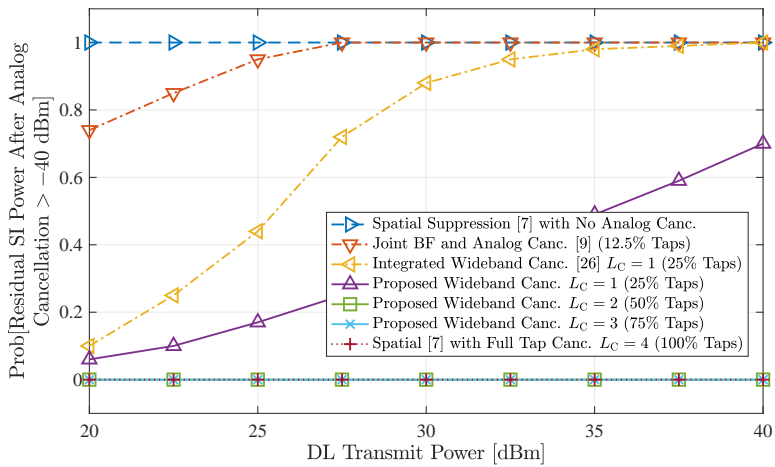}
\caption{Probability of residual SI after analog cancellation above RF saturation level $-40$ dBm with respect to DL transmit power in dBm and $N_{{\rm RX},m_1} = N_{{\rm TX},m_2}=4$.}
\label{fig: probSat}
\end{figure}

In Fig.~\ref{fig: probSat}, we plot the probability of FD MIMO node $b$ RX RF saturation for multi-antenna DL and UL users with $N_{{\rm RX},m_1} = N_{{\rm TX},m_2}=4$. With the increment of DL RX antennas, the SI effect at the node $b$ receiver grows substantially, as the TX transmits multiple streams simultaneously. For the considered multi-antenna case, the existing ``Spatial Suppression" with no analog cancellation, ``Joint BF and Analog Cancellation," and ``Integrated Wideband Cancellation" approaches provide inadequate SI cancellation performance rendering the RX RF chains into saturation. However, the proposed wideband cancellation with $50\%$ taps is still capable of achieving enough SI suppression to avoid RX RF saturation. Therefore, the proposed wideband analog SI cancellation method with $50\%$ fewer taps compared to the full-tap cancellers provides sufficient analog SI suppression avoiding RX RF chain saturation for considered single and multi-antenna users.
\begin{figure}[!tbp]
\centering
\includegraphics[width=0.7\textwidth]{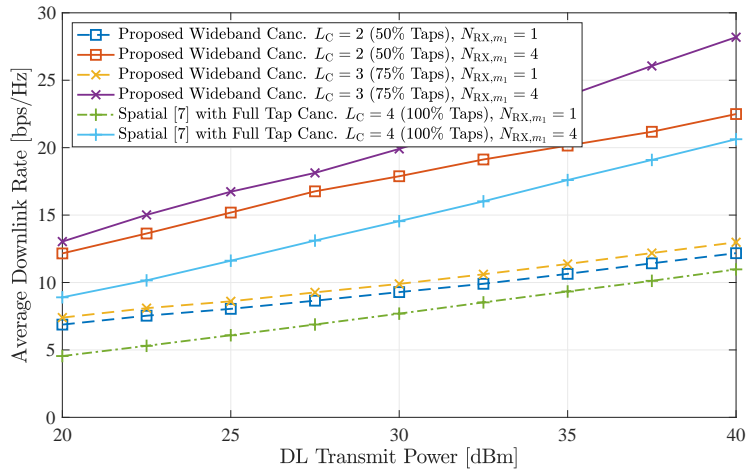}
\caption{Average downlink rate as a function of the downlink transmit power for $N_{{\rm RX},b} = N_{{\rm TX},b}=4$ and $N_{{\rm RX},m_1} = \{1,4\}$.}
\label{fig: DL_rate}
\end{figure}
\subsection{Achievable Downlink Rate}
Figure~\ref{fig: DL_rate} depicts the achievable DL rate of the proposed wideband FD MIMO system with $N_{{\rm RX},b} = N_{{\rm TX},b}=4$ and $N_{{\rm RX},m_1} = \{1,4\}$ as a function of DL transmit power. The DL rate is achieved after solving the $\mathcal{OP}2$ using Algorithm 1 given the number of analog cancellation taps that can achieve sufficient SI suppression, as discussed in the previous section. We also plot the DL rate performance of the ``Spatial suppression" approach with a full-tap analog canceller. It is evident from Fig.~\ref{fig: DL_rate} that the proposed wideband analog cancellation approach with only $50\%$ taps ($L_{\rm C} = 2$) outperforms the existing full-tap approach for both single and multi-antenna user cases. Although the proposed approach with $75\%$ taps ($L_{\rm C} = 3$) provides a substantial DL rate improvement compared to the scheme with $50\%$ taps for multi-antenna user cases, their achievable DL rate is comparable for single-antenna users. Therefore, the proposed wideband cancellation approach provides a trade-off between the number of canceller taps and the achievable DL rate. It is to be noted that, in practical implementation, it is more desirable to reduce the analog canceller taps as much as possible, providing a cost-effective solution.
\subsection{Performance of the Digital SI Canceller}
The residual SI signal after analog SI cancellation is further suppressed in the digital domain. Here, we evaluate the performance of the proposed TSVD-based adaptive digital SI cancellation approach. As discussed before, the number of parameters to be estimated for the digital SI regeneration and cancellation increases with the number of antennas and multipath SI components at node $b$. To estimate such a large number of parameters, a very high number of training samples is required resulting in an increased training overhead.

\begin{figure}[!tbp]
\centering
\includegraphics[width=0.7\textwidth]{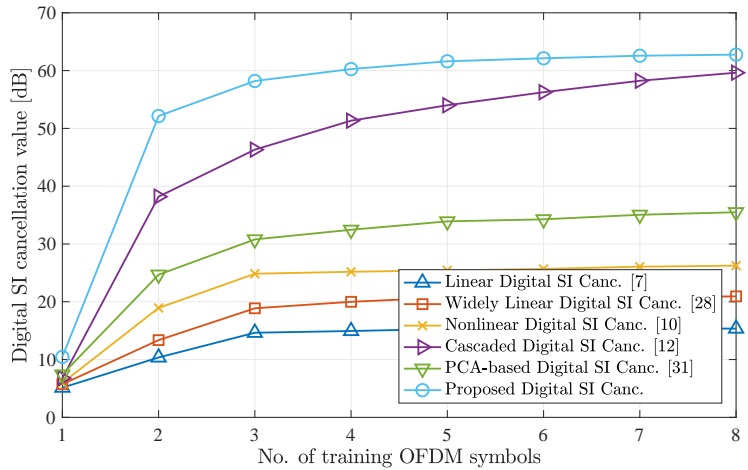}
\caption{Digital SI cancellation value with respect to the number of training OFDM symbols for digital SI estimation.}
\label{fig: DigCan}
\end{figure}
In Fig.~\ref{fig: DigCan}, we illustrate the amount of digital SI cancellation in dB with respect to the training OFDM symbols each including $64$ subcarrier samples for the considered $4\times 4$ FD MIMO node $b$ and multi-antenna users with $N_{{\rm RX},b} = N_{{\rm TX},b}= N_{{\rm RX},m_1} = N_{{\rm TX},m_2}=4$. The digital SI estimation and cancellation are achieved for a transmit power of $40$ dBm after applying the proposed wideband analog cancellation with $50\%$ taps shown to avoid RF saturation. We compare our approach with ``Linear Digital Cancellation" including only linear SI components \cite{riihonen2011mitigation}, ``Widely Linear Cancellation" considering linear and image effects \cite{korpi2014widely}, ``Nonlinear Digital Cancellation" considering third-order nonlinear SI terms \cite{bharadia2014full}, ``Cascaded Nonlinear Cancellation" providing joint cancellation technique cascading PA nonlinearity with transmitter IQ imbalance \cite{anttila2014modeling}, and ``PCA-based Nonlinear Cancellation" employing PCA transformation on SI estimation matrix \cite{korpi2017nonlinear}. It is evident that the linear, nonlinear, and widely linear cancellers achieve less than $30$ dB SI cancellation, as they only consider parts of the significant SI components for digital cancellation. The PCA-based nonlinear canceller provides around $35$ dB SI cancellation, as it omits some significant SI components to reduce estimating parameters. However, the proposed TSVD-based digital SI canceller achieves around $60$ dB cancellation value with $4$ OFDM symbols, while the cascaded nonlinear canceller requires more than double training symbols to provide a similar cancellation value. Therefore, the proposed TSVD-based digital canceller is capable of providing superior SI suppression performance with less computational resources, reducing the communication overhead.
\begin{figure}[!tbp]
\centering
\includegraphics[width=0.7\textwidth]{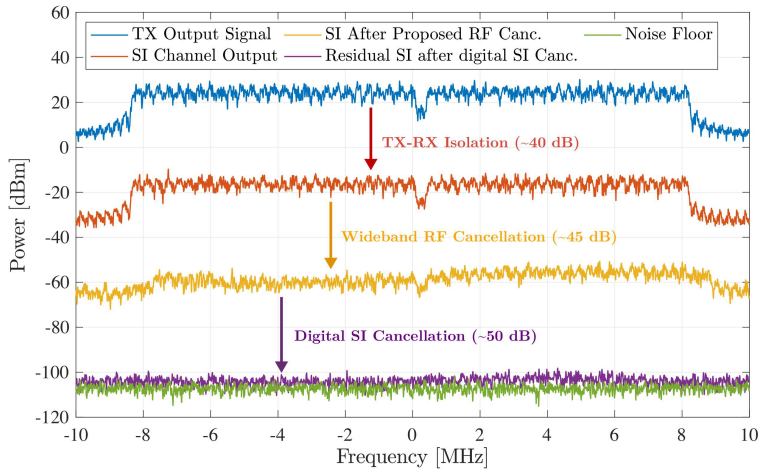}
\caption{Power Spectrum of the SI signal after SI cancellation in the RF
and digital domains with $40$ dBm average TX power, $20$ MHz bandwidth, and $-100$ dBm receiver noise floor.}
\label{fig: PowerSpectrum}
\end{figure}

Figure~\ref{fig: PowerSpectrum} depicts the power spectrum of the SI signal after TX-RX isolation, proposed wideband RF cancellation with $50\%$ taps, and TSVD-based digital SI cancellation for $40$ dBm UL and DL transmit power and $20$ MHz bandwidth with $N_{{\rm RX},b} = N_{{\rm TX},b}= N_{{\rm RX},m_1} = N_{{\rm TX},m_2}=4$. It is shown that the combination of proposed wideband A/D SI cancellation approaches can achieve around $95$ dB SI cancellation in addition to the $40$ dB TX-RX isolation, and therefore successfully suppress the SI signal to the thermal noise floor for the wideband FD MIMO systems.

\begin{figure}[!tbp]
\centering
\includegraphics[width=0.7\textwidth]{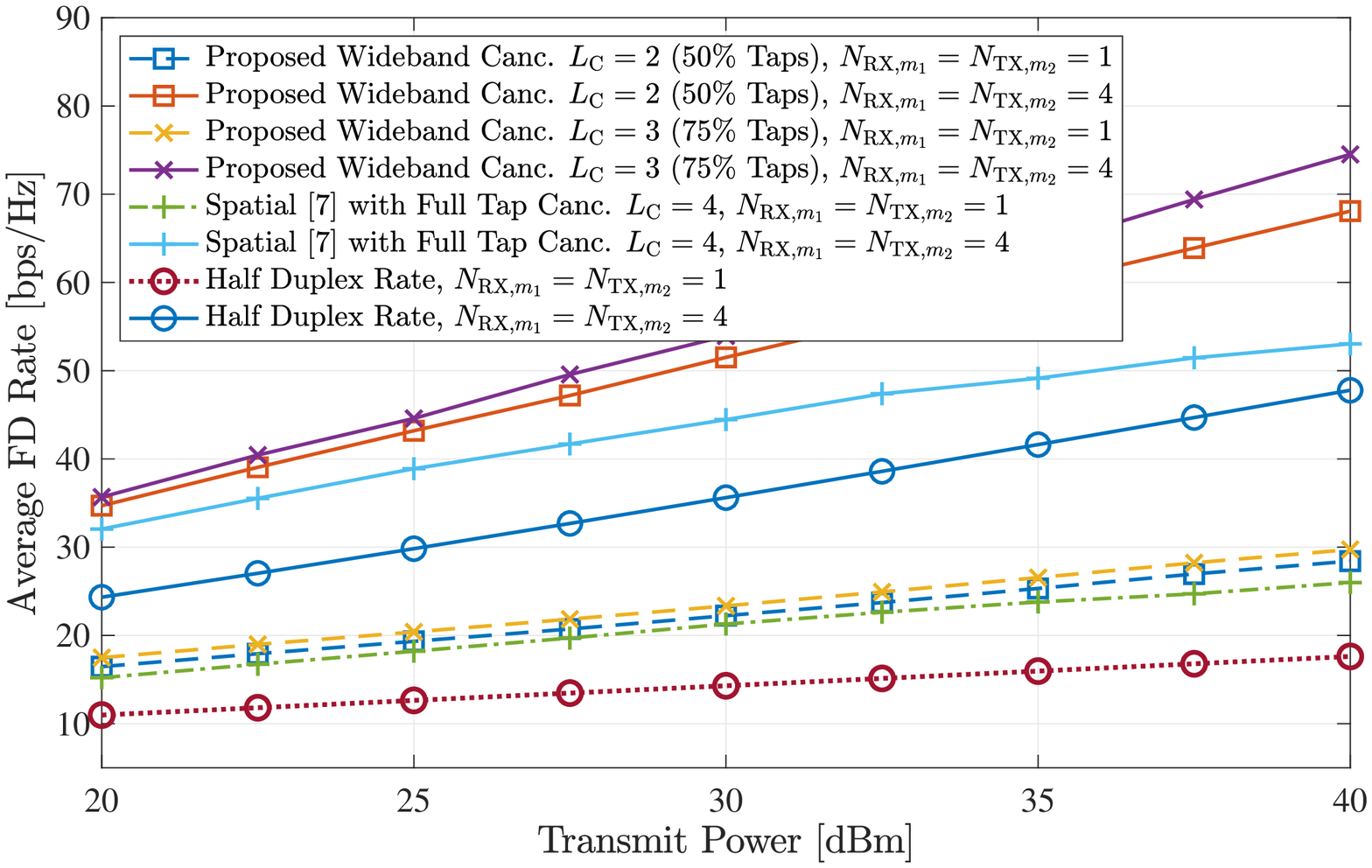}
\caption{Average FD rate as a function of the downlink transmit power for $N_{{\rm RX},b} = N_{{\rm TX},b}=4$ and $N_{{\rm RX},m_1} = N_{{\rm TX},m_2} = \{1,4\}$.}
\label{fig: FD_rate}
\end{figure}

\subsection{Average FD Rate}
\begin{figure}[!tbp]
\centering
\includegraphics[width=0.7\textwidth]{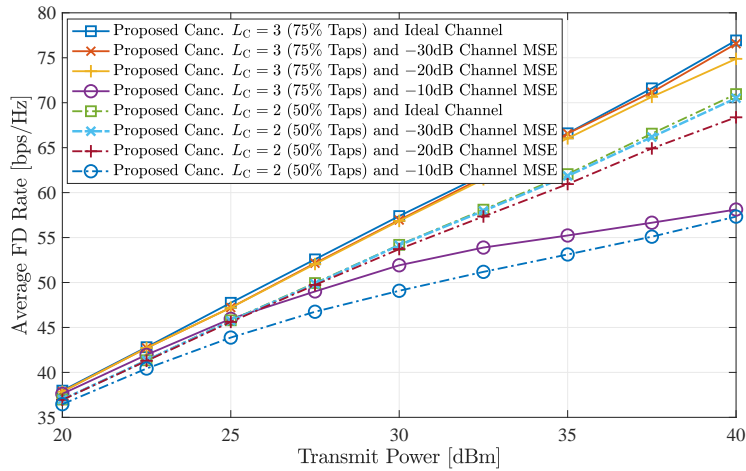}
\caption{Average FD rate as a function of the transmit power and different channel estimation MSE for $N_{{\rm RX},b} = N_{{\rm TX},b}=4$ and $N_{{\rm RX},m_1} = N_{{\rm TX},m_2} = \{1,4\}$.}
\label{fig: ChanMSE}
\end{figure}

In Fig.~\ref{fig: FD_rate}, we illustrate the FD rate of the proposed wideband A/D cancellation with respect to the DL and UL transmit power for $N_{{\rm RX},b} = N_{{\rm TX},b}=4$ and $N_{{\rm RX},m_1} = N_{{\rm TX},m_2} = \{1,4\}$. For comparison, we also provide the FD rate for ``Spatial Suppression" with full-tap analog canceller and the achievable HD rate for the considered MIMO system. It is evident that the proposed digital and analog SI cancellation approach with $50\%$ taps ($L_{\rm C} = 2$) outperforms the spatial suppression with full-tap analog canceller approach for both single and multi-antenna users cases. In addition, the proposed wideband A/D SI cancellation approach with $L_{\rm C} = 2$ can provide substantial rate improvement compared to the HD communication systems. Specifically, for a transmit power of $40$ dBm and multi-antenna users case, the proposed wideband A/D SI cancellation approach can achieve $1.45\times$ the FD rate compared to the HD system with a reduction of $50\%$ analog taps compared to the state-of-the-art full-tap canceller. 
{
\subsection{Impact of Channel Estimation Error on the FD Rate}
The impact of channel estimation error on the proposed FD system performance is depicted in Fig.~\ref{fig: ChanMSE}. We have plotted the average FD rate of the proposed wideband cancellation system with respect to the DL transmit power for $N_{{\rm RX},b} = N_{{\rm TX},b}=4$ and $N_{{\rm RX},m_1} = N_{{\rm TX},m_2} = \{1,4\}$ with different Mean Squared Errors (MSEs) for channel estimation of all considered channels (i.e. SI, DL, and UL) as well as analog cancellation taps. It is evident from the figure that, for channel estimation MSEs of $-30$dB and $-20$dB, the average FD rates are consistent with the FD rate of ideal channel knowledge for the proposed wideband A/D SI cancellation approach with both $50\%$ and $75\%$ taps. However, for the higher channel estimation error of $-10$dB MSE, the FD rate is reduced at the large transmit powers of the proposed wideband FD system. This happens due to the utilization of increased DL DoFs in order to achieve the required SI cancellation, which will avoid RF saturation with non-ideal channels.
}

\begin{figure}[!tbp]
\centering
\includegraphics[width=0.7\textwidth]{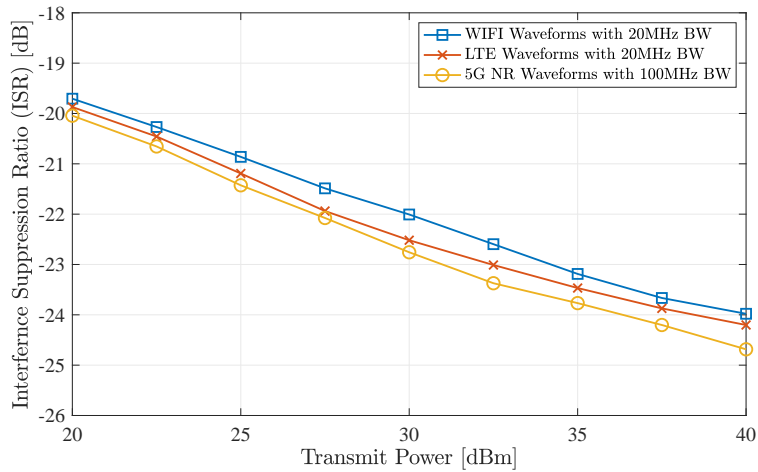}
\caption{Interference Suppression Ratio with respect to the transmit power and different signal waveforms for the considered FD MIMO system with $N_{{\rm RX},b} = N_{{\rm TX},b}= N_{{\rm RX},m_1} = N_{{\rm TX},m_2} =4$.}
\label{fig: ISR}
\end{figure}

{
\subsection{Interference Suppression Ratio for Different Signal Waveforms}
In Fig.~\ref{fig: ISR}, we illustrate Interference Suppression Ratio (ISR), which is defined as the average residual interference power versus the average SI power without cancellation, with respect to the transmit power for $N_{{\rm RX},b} = N_{{\rm TX},b}= N_{{\rm RX},m_1} = N_{{\rm TX},m_2} =4$ and different signal waveforms. In addition to the WiFi signal waveform with $20$MHz BW and $312.5$KHz subcarrier spacing as shown in Table~\ref{tab: sim_param}, we have also simulated Long Term Evolution (LTE) waveform with $20$MHz BW and $15$KHz subcarrier spacing as well as 5G New Radio (NR) waveform with $100$MHz BW and $60$KHz subcarrier spacing, while considering multipath SI channel with $L_{\rm C} = 4$ as described in Sec.~\ref{subsec:sim_param}. It is to be noted that 5G NR and LTE waveforms' sample delays, which are $4.069$ns and $32.552$ns corresponding to the $60$KHz and $15$KHz subcarrier spacing, respectively, are not similar to the SI path delays ($50$ns delay between consecutive reflected paths). Therefore, for such waveforms, analog canceller experiences larger phase error corresponding to larger delay mismatch error. However, it is evident in Fig.~\ref{fig: ISR} that the proposed joint wideband A/D SI cancellation and TX/RX beamforming approach can achieve similar ISR performance for $100$MHz 5G NR and $20$MHz LTE waveforms to the previously considered WiFi waveforms. Therefore, the proposed approach is capable of achieving the required SI cancellation for both LTE and 5G NR waveforms with large bandwidth.
}

\section{Conclusion}\label{sec: conclusion}
In this paper, we have presented a reduced complexity wideband analog SI canceller jointly designed with digital TX/RX beamforming for practical FD MIMO systems with TX IQ imbalances and PA nonlinearities maximizing the FD sum-rate performance. To suppress the residual linear SI signal along with its conjugate and nonlinear components below the noise floor, we proposed a novel adaptive digital SI cancellation technique reducing the number of estimation parameters. Our representative performance evaluation comparisons with existing wideband FD MIMO designs demonstrated that the proposed wideband A/D SI canceller achieves superior SI cancellation capability. Furthermore, we showed that the proposed optimization framework provided higher achievable rate performance with reduced hardware complexity for analog SI cancellation and computational resources of digital SI cancellation compared to the full-tap wideband FD MIMO radios. For future research, we intend to extend the proposed hardware and algorithmic co-design framework to massive MIMO FD radios with hybrid A/D beamformers, considering multiple users at both the UL and DL directions.
\newpage

\section*{Appendix A\\Detailed Derivation of \eqref{eq: est_s_b}}
The linearly processed estimated symbol vector $\widehat{\mathpzc{s}}_{b,n}$ is derived as
\begin{align*}
        \widehat{\mathpzc{s}}_{b,n} = &\; 
        \mathbfcal{U}_{m_1,n} \left( \frac{1}{\sqrt{N_c}}\sum\limits_{k=0}^{N_c-1}\left(\sum\limits_{\ell=0}^{L_{\rm DL}-1} \H_{\rm DL}[\ell]\widetilde{\x}_b[k-\ell] +  \w_{m_1}[k]\right)e^{-\frac{j2\pi kn}{N_c}}\right)\\
        \stackrel{(a)}{=}&\; \mathbfcal{U}_{m_1,n} \left( \frac{1}{\sqrt{N_c}}\sum\limits_{k=0}^{N_c-1}\left(\sum\limits_{\ell=0}^{L_{\rm DL}-1} \H_{\rm DL}[\ell]\left(\G_{1,b}\x_b[k-\ell] + \z[k-\ell]\right) + \w_{m_1}[k]\right)e^{-\frac{j2\pi kn}{N_c}}\right)\\
        \stackrel{(b)}{=}&\; \mathbfcal{U}_{m_1,n} \Bigg(\frac{1}{\sqrt{N_c}}\sum\limits_{k=0}^{N_c-1}\Bigg(\sum\limits_{\ell=0}^{L_{\rm DL}-1}\H_{\rm DL}[\ell]\bigg(\G_{1,b}\frac{1}{\sqrt{N_c}} \sum\limits_{p=0}^{N_c-1} \mathbfcal{V}_{b,p}\mathpzc{s}_{b,p} e^{\frac{j2\pi p(k-\ell)}{N_c}} + \z[k-\ell] \bigg) \\&+ \w_{m_1}[k]
        \Bigg)  e^{-\frac{j2\pi kn}{N_c}}\Bigg) \\
        % &\;  + \frac{1}{\sqrt{N_c}}\sum\limits_{k=0}^{N_c-1}\w_{m_1}[k]e^{-\frac{j2\pi kn}{N_c}}
        % \Bigg)\\
        =&\; \mathbfcal{U}_{m_1,n} \Bigg(\sum\limits_{\ell=0}^{L_{\rm DL}-1}\H_{\rm DL}[\ell]e^{-\frac{j2\pi \ell n}{N_c}}\Bigg(\G_{1,b}\bigg(\frac{1}{{N_c}} \sum\limits_{k=0}^{N_c-1}\sum\limits_{p=0}^{N_c-1} \mathbfcal{V}_{b,p}\mathpzc{s}_{b,p} e^{\frac{j2\pi (p-n)k}{N_c}}e^{-\frac{j2\pi \ell p}{N_c}}\bigg) \\
        &\;
        + \frac{1}{\sqrt{N_c}}\sum\limits_{k=0}^{N_c-1}\z[k-\ell]e^{-\frac{j2\pi (k-\ell)n}{N_c}}\Bigg) +
        \frac{1}{\sqrt{N_c}}\sum\limits_{k=0}^{N_c-1}\w_{m_1}[k]e^{-\frac{j2\pi kn}{N_c}}
        \Bigg)\\
        \stackrel{(c)}{=}&\; \mathbfcal{U}_{m_1,n}\Bigg(\mathbfcal{H}_{{\rm DL},n}\left(\G_{1,b}\mathbfcal{V}_{b,n}{\mathpzc{s}}_{b,n} + {\mathpzc{z}}_{b,n}\right) + \mathpzc{w}_{m_1,n}
        \Bigg),
\end{align*}
where $(a)$ and $(b)$ are obtained from \eqref{eq: signal_DL} and \eqref{eq: ifft_b}, respectively. Here, $(c)$ is derived using the identities $\frac{1}{{N_c}}\sum\limits_{k=0}^{N_c-1}e^{\frac{j2\pi (p-n)k}{N_c}} = \bigg\{
        \begin{array}{c c}
            1, & \text{if}\,\, p = n,\\
            0, & \text{if}\,\, p \neq n,
        \end{array}$, $\mathpzc{z}_{b,n}= \frac{1}{\sqrt{N_c}}\sum\limits_{k=0}^{N_c-1}\z[k]e^{-\frac{j2\pi k n}{N_c}}$, and  $\mathpzc{w}_{m_1,n}=\frac{1}{\sqrt{N_c}}\sum\limits_{k=0}^{N_c-1}\w_{m_1}[k]e^{-\frac{j2\pi kn}{N_c}}$.\\

Similarly, the proof of \eqref{eq: fftOutput_b} can be obtained using \eqref{eq: freq_rep} and the above mentioned identities.

\bibliographystyle{IEEEtran}
\bibliography{mybib.bib}

% \begin{IEEEbiography}{Md Atiqul Islam}
% Biography text here.
% \end{IEEEbiography}

% % if you will not have a photo at all:
% \begin{IEEEbiography}{George C. Alexandropoulos}
% Biography text here.
% \end{IEEEbiography}

% \begin{IEEEbiography}{Besma Smida}
% Biography text here.
% \end{IEEEbiography}

\end{document}